\documentclass[3p,times]{elsarticle}
\pdfoutput=1

\usepackage[version=4]{mhchem} 
\usepackage{subcaption}
\usepackage{multirow}
\usepackage{siunitx}
\usepackage{textcomp}
\usepackage{comment}
\usepackage{color}
\usepackage{tikz}
\usepackage{placeins}
\usepackage{siunitx}
\usepackage{wrapfig}
\makeatletter
\def\ps@pprintTitle{%
 \let\@oddhead\@empty
 \let\@evenhead\@empty
 \def\@oddfoot{\centerline{\thepage}}%
 \let\@evenfoot\@oddfoot}
\makeatother

\begin{document}

\begin{frontmatter}

\title{Temperature Dependence of Self-diffusion in \ce{Cr2O3} from First Principles}

    \author[1,2]{Bharat Medasani}
    \author[1]{Maria L. Sushko}
    \author[1]{Kevin M. Rosso}
    \author[3]{Daniel K. Schreiber}
    \author[3]{Stephen M. Bruemmer}
    \cortext[author]{Corresponding author: B. Medasani (mbkumar@udel.edu, mbkumar@gmail.com). }

    \address[1]{Physical and Computational Sciences Directorate, Pacific Northwest National Laboratory, Richland, WA 99354, USA }
    \address[1]{Delaware Energy Institute, University of Delaware, Newark, DE 19702, USA }
    \address[3]{Energy and Environment Directorate, Pacific Northwest National Laboratory, Richland, WA 99354, USA }
\begin{abstract}
Understanding and predicting the dominant diffusion processes  in \ce{Cr2O3} is essential to its optimization for anti-corrosion coatings, spintronics, and other applications. Despite significant theoretical effort in modeling defect mediated diffusion in \ce{Cr2O3} the correlation with experimentally measured diffusivities remains poor partly due to the insufficient accuracy of the theoretical approaches. Here an attempt to resolve these discrepancies is made through high accuracy density functional theory simulations coupled with grand canonical formalism of defect thermochemistry.  In this approach, point defect formation energies were computed using hybrid exchange correlation functional. This level of theory proved to be essential for achieving the agreement with experimental self-diffusion coefficients.  The analysis of the resulting self-diffusion coefficients indicate that chromium has higher mobility at low temperatures 
and high oxygen partial pressures, in particular at standard temperature and pressure conditions. At high vacuum, high temperature conditions, oxygen diffusion becomes dominant. At Cr/\ce{Cr2O3} interfaces O vacancies were
found to be more mobile than Cr vacancies at all temperatures. Cr diffuses preferentially  along the c-axis at low temperatures but switches to basal plane at higher temperatures. O diffusion is primarily bound to basal plane at all temperatures. 
\end{abstract}

\end{frontmatter}

\section{Introduction}
\ce{Cr2O3} finds applications in both functional and structural applications such as catalysis~\cite{bates2015,xie2013}, electro-optics\cite{arca2017,farrell2015}, magnetoelectronic~\cite{mu2013, pati2015}, and corrosion 
protection\cite{Zurek2008}. It serves as a protective oxide coating on   widely used high-temperature Ni and Fe alloys\cite{schreiber2013,Schreiber2014,brozek2016}. 
Point defects play an important role in the effectiveness of \ce{Cr2O3} for corrosion protection and other
applications and significant experimental effort has been devoted to  understanding the nature of point defects and
their diffusion mechanisms. However, there is a large spread in the reported self-diffusion coefficients and 
 defect formation energies 
from various experimental studies~\cite{Sabioni1992,sabioni1992a,sabioni1992b,sabioni1992c,laturomain2017,Huntz1994,kofstad1982,tsai1996,hoshino1983,Schmucker2016}.

Complementing the experimental investigations, diffusion mediated by both intrinsic and extrinsic point defects in \ce{Cr2O3} have been 
extensively studied theoretically by many groups, including ours~\cite{catlow1988,Lebreau2014,medasani2017,gray2016,cao2017,Vaari2015,rak2018,medasani2018}. Due to the advances in  density functional theory
(DFT) accuracy in calculating many properties of materials, recent studies employed DFT to evaluate  formation energies, migration barriers and charge localization of point defects\cite{Lebreau2014,medasani2017,gray2016,medasani2018,rak2018}. 
The 0K energies obtained from DFT were combined with Einstein-Smoluchowski random walk thermal diffusion formalism\cite{mehrer2007} to predict the self-diffusion coefficients in \ce{Cr2O3}. However, the agreement between calculated and experimental self-diffusion coefficients had, so far, been poor, which prompted the use of modified diffusion equations~\cite{rak2018,Lebreau2014}. This leaves the question of the root cause of the discrepancy between theory and experiment unresolved. 

A possible source of discrepancy is the insufficient accuracy of the DFT in modeling transition metal oxides, which requires the use of empirical corrections to the exchange-correlation functionals to induce electron localization. 
For example, to correct for the self-interaction error and induce electron localization in modeling \ce{Cr2O3} highly correlated Cr-3d (and sometimes O-2p) electrons the Hubbard on-site correction is often employed~\cite{dftu_liech,dftu_dudarev,Rohrbach2004}. However, the resulting defect formation energies vary significantly~\cite{Lebreau2014,medasani2017,gray2016,cao2017,Schmucker2016,kehoe2016} 
depending on the choice of Hubbard U parameters and the type of the semi-local generalized 
gradient approximation (GGA) functionals.  Adding to the confusion, corrections to minimize the electrostatic errors inherent in the periodic supercell formulation of defects and the errors due to the under-prediction of bandgap by semi-local functionals  were either not applied or inconsistently applied in many of these studies. 

Rak and Brenner found that the  barrier energies for defect migration were not sensitive to the variations in U value. In contrast, they showed that varying U value results in a variation of defect formation energies~\cite{rak2018}. The likely reason for such variations is in electronic dissimilarities in the local  environments in defected and bulk systems. For example, the electronic density surrounding a point defect such as a vacancy is much lower compared to the electronic density in the corresponding bulk system and the relative effect of U is different.

Selecting the U parameter in the Hubbard correction scheme is neither trivial nor consistent. 
Often, U selection is based on fitting a bulk property to the experimental value. The chosen optimal U value 
is highly dependent on the selected bulk property. Another approach which avoids the empirical route in the selection of U is based on the linear response theory (LRT)\cite{Cococcioni2005}. LRT based U values often tend to be smaller than the U values obtained from  regressing  bulk properties. To overcome the uncertainties associated with the empirical U value of
Hubbard correction scheme, we employed hybrid functional to 
accurately evaluate defect electronic levels and defect formation energies. 

Here we report self-diffusion coefficients computed using Einstein-Smoluchowski formalism
and high accuracy point defect energies. The resulting diffusion coefficients have a reasonable agreement with the experimental data. 
This work can be considered as the culmination of our previous studies on the atomistic diffusion mechanisms
in \ce{Cr2O3} mediated by vacancies and interstitials~\cite{medasani2017,medasani2018}.
For barrier energies, the transition state and ground defect state exhibit  similar local electronic environment, suggesting an insensitivity to the U value used in this prior work. Hence, by utilizing the transition barrier energies obtained in those studies in addition to the revised more accurate defect energies and defect thermodynamics, we evaluated
Fermi level variations, stoichiometry deviations and self diffusion coefficients in undoped
bulk \ce{Cr2O3} as a function of temperature and oxygen partial pressure.  The computed
self diffusion coefficients were analyzed using Brouwer diagrams to reveal the dominant defects responsible for diffusion.

\section{Methods\label{sec:methods}}
\subsection{Density Functional Theory Approach}
Density functional theory (DFT) calculations for defect formation energies were performed using the 
Vienna \textit{ab initio} simulation package (VASP)~\cite{kresse93,kresse94,kresse96}. 
We utilized HSE~\cite{hse06} hybrid functional with 25\% mixing of Hartree - Fock exchange at short range and 100\% PBE correlation. Projector augmented wave (PAW)~\cite{blochl94,kresse99} optimized
for PBE~\cite{pbe} functional was used. Wave functions preoptimized from GGA+U calculations with PBEsol~\cite{pbesol} functional 
was used as input to the HSE calculations. The HSE screening parameter of 0.6, which is different from the commonly used values of 0.2 and 0.3 corresponding to HSE06 and HSE03, respectively, was chosen after fitting the computed bandgaps with the experimental bandgap. This point is further discussed in the Results section.
The reciprocal space of the primitive cell of \ce{Cr2O3} ($R\overline{3}c$ space group) with 10 atoms was sampled  with 5$\times5$$\times5$ $\Gamma$-centered \textit{k}-point grid. The primitive cell was fully relaxed (both size and shape were relaxed) until the forces converged to 0.01 eV/\AA. 
The atomic positions in the 2$\times2$$\times1$ defect supercells were relaxed at constant volume 
and fixed cell shape 
until the individual forces on each atom were minimized to 0.03 eV/\AA. 
For supercell calculations, a cut-off value of 400 eV was used for the plane-wave basis set and 2$\times2$$\times1$ 
$\Gamma$-centered \textit{k}-point grid was used to sample the reciprocal space.
 Spin polarization with anti-ferromagnetic (AFM) ordering of 
Cr spins was used. Gaussian method with a width of 0.01 eV was used for 
electronic smearing.

\subsection{Diffusion Coefficient Calculation}
In materials which support charged defects, the diffusion coefficient of an element $s$ can be defined as
\begin{equation}
D_s = \sum_X \sum_q d(X_s^q)
\label{eq:diff_coeff_sum}
\end{equation}
where $d$ is the diffusion coefficient pertaining to a defect $X_s^q$, with $X$ denoting 
the defect type, and $q$ representing the defect charge. X can be a regular point defect, such
as a vacancy or an interstitial, or it can be a complex defect, such as the Frenkel defect. Based on Einstein's random walk theory,  $d$ can be computed as 
\begin{equation}
d = \frac{1}{2}c\sum_p{m_p l_p ^ 2 \Gamma_p}.
\label{eq:diff_coeff}
\end{equation}
Here $c$ is the concentration of defect, $X_s^q$, and $p$ represents one of the migration 
pathways. $m_p$, $l_p$, and $\Gamma_p$ indicate the multiplicity, length, 
and jump  frequency associated with path $p$, respectively.  The defect concentration, $c$, is a function of 
the defect formation energy $E_f$,
\begin{equation}
c = c_0  e^{-\beta E_f},
\label{eq:def_conc}
\end{equation}
were $c_0$ is the number of lattice sites per cell volume.
According to Vineyard's transition state theory (TST)~\cite{vineyard1957}, the jump frequency 
is defined as a function of the attempt frequency $\nu_p$ and the migration barrier energy $E_{m,p}$ as
\begin{equation}
\Gamma_p = \nu_p e^{-\beta E_{m,p}}.
\label{eq:jump_freq}
\end{equation}
The attempt frequency is evaluated under harmonic approximation using the phonon modes of defect ground and 
transition states as
\begin{equation}
\nu = \frac{\prod_i^{3N-3} {\nu_i}} { \prod_i^{3N-4} {\nu'_j}},
\end{equation}
where $\nu_i$ and $\nu'_j$ represent the real phonon frequencies of  the ground and transition states
of the defect respectively and N is the number of atoms in the bulk supercell.

\section{Finite Temperature Defect Concentrations}
The effect of  pressure and temperature is accounted for in the diffusion coefficients 
through the chemical potential term in the expression for defect formation energies. 
 Assuming dilute concentrations for defects, we utilized constrained grand canonical  
 formalism to compute the elemental chemical 
 potentials at finite temperatures. This formalism has been successfully applied to predict 
 defect concentrations in \ce{ZrO2} under dilute conditions\cite{youssef2012}. In this formalism, bulk 
 \ce{Cr2O3} is assumed to be in contact with an infinite \ce{O2} reservoir at temperature T and partial 
pressure $p_{O_2}$. The oxygen chemical potential then is defined as
\begin{equation}
\mu_O(p_{O_2}, T) = \mu_O^0  +  \Delta\mu_O^0(1 atm, 298 K)  + \Delta\mu_O(p_{O_2}, T).
\label{cgc_o_chem}
\end{equation}
In the above equation, $\mu_O^0$ is the 0 K DFT computed O chemical potential, $\Delta\mu_O^0(1\,atm,\,298\,K)$ 
is the correction to obtain the experimental chemical potential of O at standard temperature and pressure (STP: 273 K and 1 atm pressure) 
conditions, and $\Delta\mu_O(p_{O_2}, T)$ is given by
\begin{equation}
\Delta\mu_O(p_{O_2}, T) = kT\, ln(p_{O_2}). 
\end{equation}
The latter two quantities are obtained from NIST-JANAF tables\cite{nist-janaf}. Cr chemical potential is obtained from $\mu_O$ as
\begin{equation}
\mu_{Cr}(p_{O_2}, T) = \Delta H_{Cr_2O_3} - \mu_O(p_{O_2}, T) , 
\end{equation}
where $\Delta H_{Cr_2O_3}$ is the formation enthalpy of \ce{Cr2O3}.

Since the overall system is charge neutral, we impose the charge neutrality condition 
after accounting for the charged defects and free carriers such as electrons and holes using the expression
\begin{equation}
\sum_{X^q_s} q\, c^{X^q_s} + p - n = 0,
\label{charge_neutral}
\end{equation}
where $p$ and $n$ represent the number densities of holes and electrons, respectively.
Given that the formation energy of a charged defect can be calculated as~\cite{medasani2017,medasani2018}
\begin{equation}
E_f^{X^q_s} = E_{tot}^{X^q_s} - E_{tot}^{bulk} + \sum_{s'}{ n_{s'}(X^q_s)\mu_{s'}} + qE_F + E_{corr}^{X^q_s}.
\label{eq:gen_def_form_en}
\end{equation}
We can rewrite the concentration of defect $X_s^q$ as
\begin{equation}
c^{X^q_s}(p_{O_2},T, E_F) = c_{0}^{X^q_s}e^{-E_f^{X_s^q}(p_{O_2}, T, E_F)/kT}.
\label{def_conc}
\end{equation}

\begin{wraptable}{l}{0.3\textwidth}
\renewcommand{\arraystretch}{1.1}
\caption{Bandgap of \ce{Cr2O3} obtained with different values of HSE functional screening parameter. }
\centering
\begin{tabular}{cc}
\hline
$\mu$ & Bandgap (eV) \\ 
\hline
0.2 & 4.27  \\
0.3 & 3.85  \\
0.4 & 3.64  \\
0.5 & 3.59  \\
0.6 & 3.38  \\
\hline
\end{tabular}
\label{tab:hse_bandgap_fit}
\end{wraptable}
The number density of free electrons, $n$, is given by 
\begin{equation}
n(T, E_F) = \int_{CBM}^\infty{f_{FD}(E,T,E_F)}g(E)dE,
\label{Eq:n}
\end{equation}
where g(E) is the density of states, $f_{FD}$ is the Fermi-Dirac distribution function given by
\begin{equation}
f_{FD}(E,T,E_F) = \frac{1}{1+e^{(E-E_F)/kT}}.
\end{equation}
Similarly the number of free holes, $p$, is given by
\begin{equation}
p(T,E_F) = \int_{-\infty}^{VBM}{(1-f_{FD}(E,T,E_F)})g(E)dE.
\label{Eq:p}
\end{equation}

\begin{wraptable}{l}{0.52\textwidth}
\renewcommand{\arraystretch}{1.1}
\caption{Bulk properties of \ce{Cr2O3} obtained with HSE functional.
}
\begin{tabular}{lccc}
\hline
Property & This Work & \multicolumn{2}{c}{Literature} \\ 
& & Theory (GGA+U) & Experiment\\
\hline
{\it a} & 4.925  & & \\
{\it c}  &13.54   & 13.68~\cite{medasani2017}, 13.85~\cite{Lebreau2014} & 13.566\\
$\mu$   & 2.81  & 2.91~\cite{medasani2017}, 3.1~\cite{Lebreau2014} & 2.8 \\
Bandgap & 3.38  & 2.95~\cite{medasani2017}, 2.8~\cite{Lebreau2014} & 3.4\\
\hline
\end{tabular}
\label{tab:hse_bulk_cr2o3}
\end{wraptable}
Substituting Eqns.~\ref{def_conc}, \ref{Eq:n}, and \ref{Eq:p} into
Eqn.~\ref{charge_neutral}, and  by making
$T$ and $p_{O_2}$ as free variables, we can solve for $E_F$ in terms of $T$ and $p_{O_2}$ self-consistently.  
The resulting defect concentrations are given by
\begin{equation}
c^{X^q_s}(p_{O_2},T) = c_{0}^{X^q_s}e^{-E_f^{X_s^q}(p_{O_2}, T)/kT}.
\label{def_conc_free}
\end{equation}

\section{Results and Discussion \label{sec:results}}
\subsection{HSE Screening Parameter}
To eliminate the uncertainty surrounding the defect formation energies computed with GGA+U formalism, we utilized HSE hybrid functional. 
The bandgap of \ce{Cr2O3} predictions by the two widely used flavors of HSE, of HSE06 and HSE03,~\cite{hse06} hybrid functional, of 4.27 eV and 3.85 eV, respectively, were much higher than the experimental values of 3.2-3.4 eV. In order to lower the calculated bandgap closer to the experimental value of 3.4 eV, we modified the parameters of the HSE functional. We retained the 25\% mixing of Hartree-Fock exchange at short range  and increased the screening parameter till the computed bandgap matched with the experimental value. For bandgap fitting, we used a fixed cell size that was optimized with PBEsol+U approach in our previous studies.~\cite{medasani2017,medasani2018} The bandgaps obtained with different screening parameters are listed in Table~\ref{tab:hse_bandgap_fit}. 

\begin{wrapfigure}{r}{0.5\textwidth}
\centering
\includegraphics[trim={0.4cm 0.3cm 0 7.6cm},clip,width=1\linewidth]{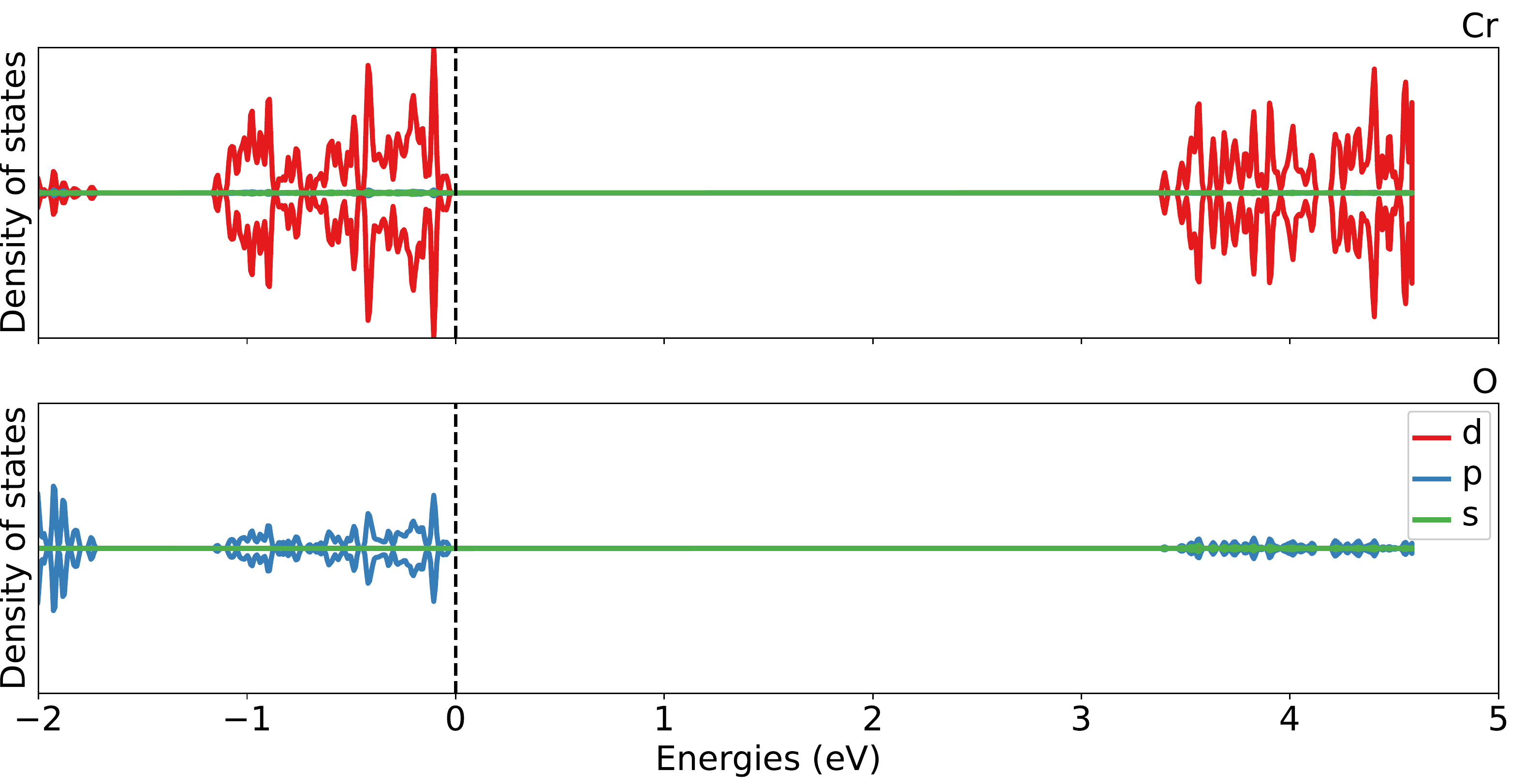}
\caption{Electronic DOS of bulk \ce{Cr2O3} computed with HSE functional with 0.6 screening parameter.} 
\label{fig:Cr2O3_bulk_dos}
\end{wrapfigure}
The results indicate that calculations with the screening parameter of 0.6 provide the best match between the computed and experimental  bandgap. 
    The primitive cell was then optimized with the chosen HSE parameters. 
 The resulting cell dimensions, magnetic moment and bandgap are given in Table~\ref{tab:hse_bulk_cr2o3}. Specifically, both \textit{a} and \textit{c} are smaller compared to those obtained using GGA+U and closer to the experimental values. Similarly, the computed magnetic moment of Cr has a better match with the experimental value. After relaxation, the bandgap increased negligibly from 3.377 eV to 3.384 eV. 
 The resulting electronic density of states (DOS) of bulk \ce{Cr2O3} is given in Figure~\ref{fig:Cr2O3_bulk_dos}.

 Cr-O phase diagram was evaluated by optimizing the Cr, \ce{O2}, and \ce{CrO2} phases with HSE parameters identical  to those used in the relaxation of  \ce{Cr2O3} cell. The resulting  phase diagram (given in SI Figure S1) yielded the formation energies of \ce{Cr2O3} and \ce{CrO2} at $-2.54$ eV/atom and $-2.17$ eV/atom, respectively. The computed formation energy of \ce{Cr2O3} compares well with the corresponding experimental value of $-2.35$ eV/atom. 
 
 \subsection{Defects}
    \begin{figure*}[!htb]
\begin{subfigure}{0.5\linewidth}
\centering
\includegraphics[trim={0.4cm 0.3cm 0 0.4cm},clip,width=1\linewidth]{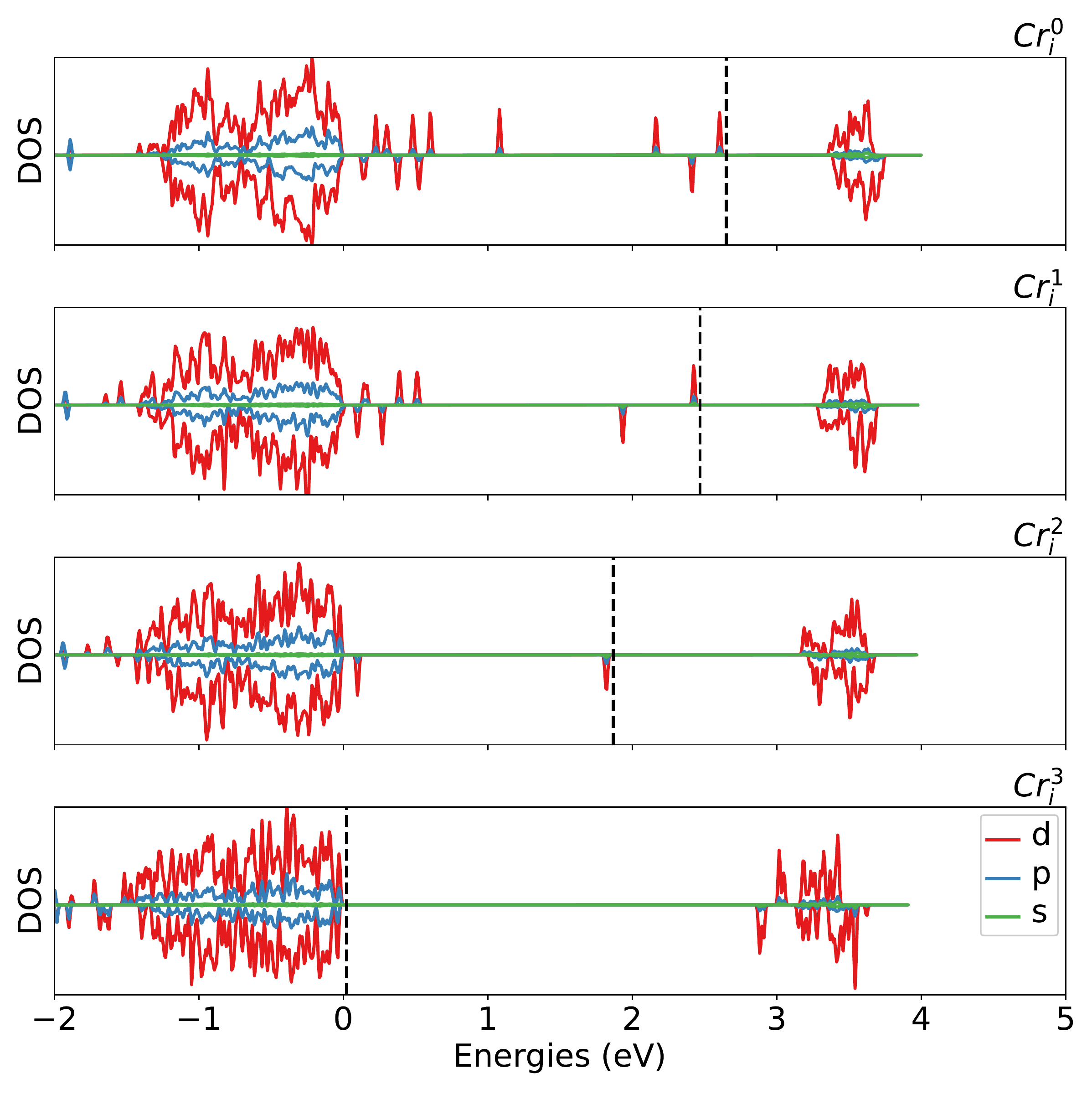}
\caption{Cr interstitials}
\label{fig:Cr_inter_dos}
\end{subfigure}%
\begin{subfigure}{0.5\linewidth}
\centering
\includegraphics[trim={0.4cm 0.3cm 0 0.4cm},clip,width=1\linewidth]{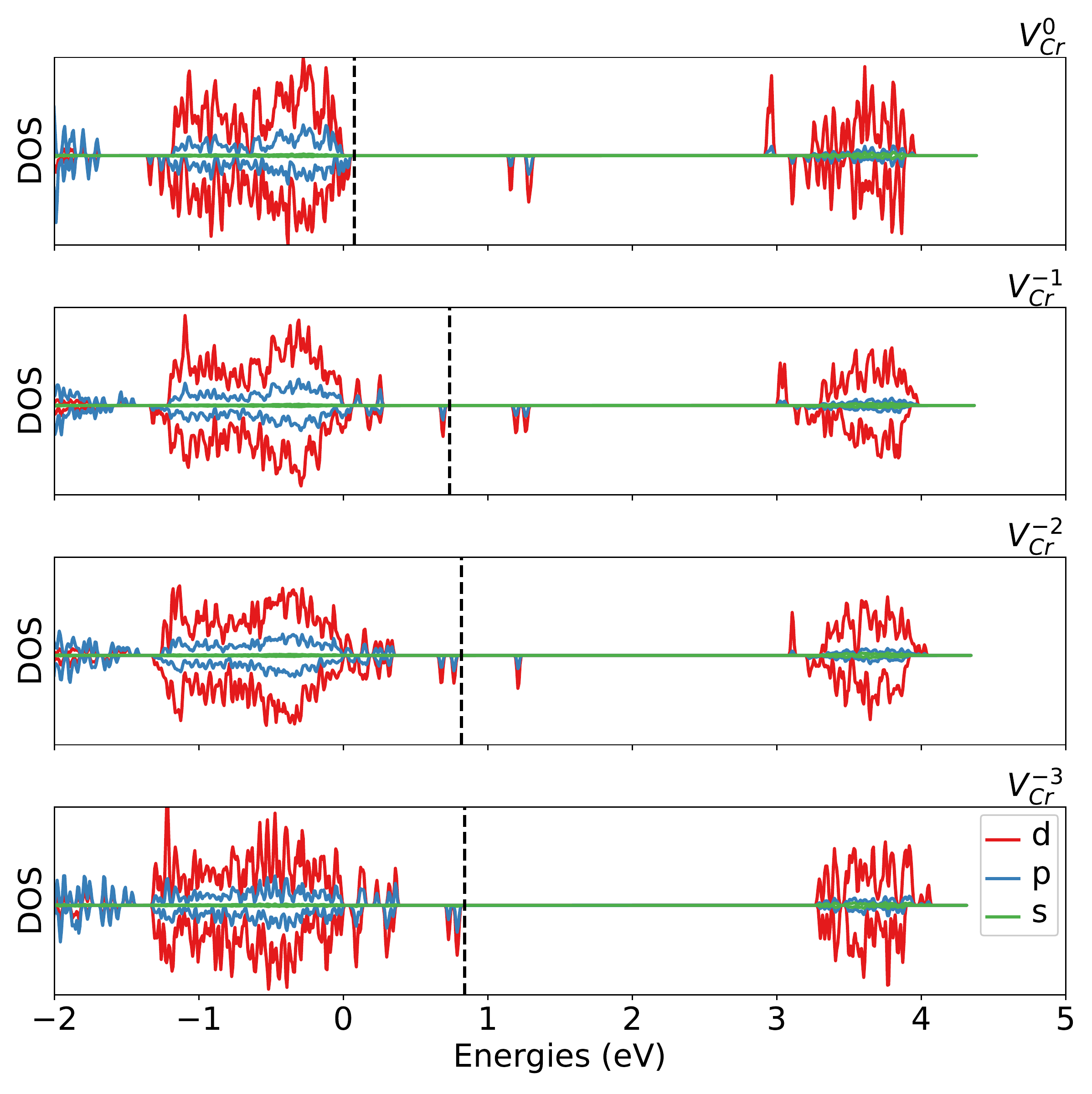}
\caption{Cr vacancies }
\label{fig:Cr_vac_dos}
\end{subfigure}
\caption{Orbital projected DOS averaged over all sites for interstitials in different charge states. Dashed vertical line marks the position of the Fermi level. The zero level has been fixed at the VBM.}
\label{fig:Cr_def_dos}
\end{figure*}
\begin{wrapfigure}{r}{0.5\linewidth}
\centering
\includegraphics[trim={0.4cm 0.3cm 0 0.4cm},clip,width=1\linewidth]{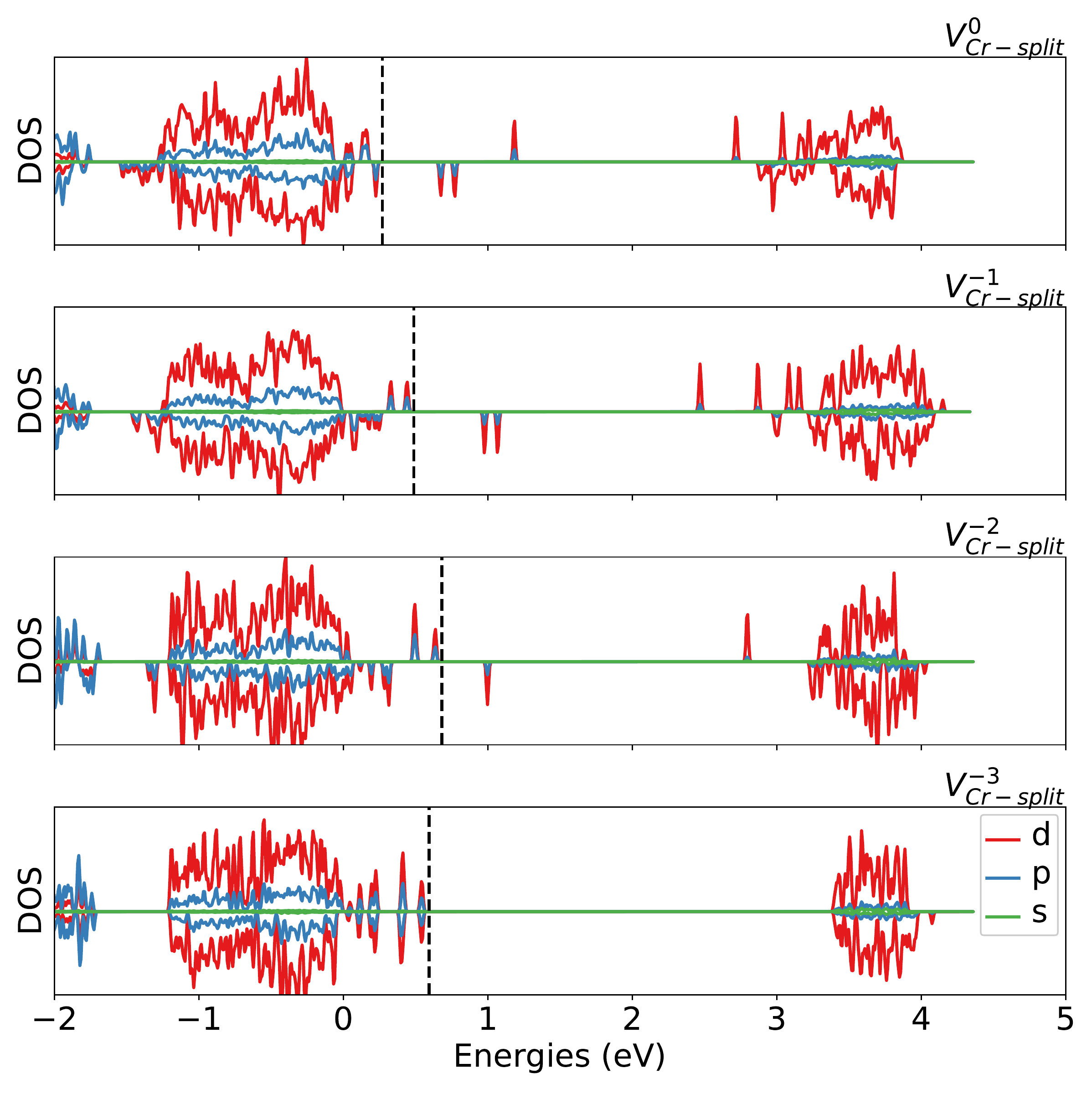}
\caption{Orbital projected DOS averaged over all sites for Cr vacancy triple defects in different charge states. Dashed vertical line marks the position of the Fermi level. The zero level has been fixed at the VBM. }
\label{fig:Cr_vac_TD_dos}
\end{wrapfigure}
Defect calculations were performed using a 2$\times2$$\times1$ supercell containing vacancies and interstitials and generated from the optimized primitive cell. In addition to vacancies and interstitials, Cr vacancy triple defect (or split-vacancy) identified in our earlier work~\cite{medasani2017} was also considered. The ionic positions in the defect supercells were optimized under fixed volume and shape conditions. 

To gain an understanding of how HSE affected defect properties, we evaluated electronic density of states (DOS),  distribution of excess electrons and holes for charged defects, and the formation energies of the optimized defects.

\subsubsection{Electronic Structure of Defects}
\begin{figure*}[!htb]
\begin{subfigure}{0.5\linewidth}
\centering
\includegraphics[trim={0.4cm 0.3cm 0 0.4cm},clip,width=1\linewidth]{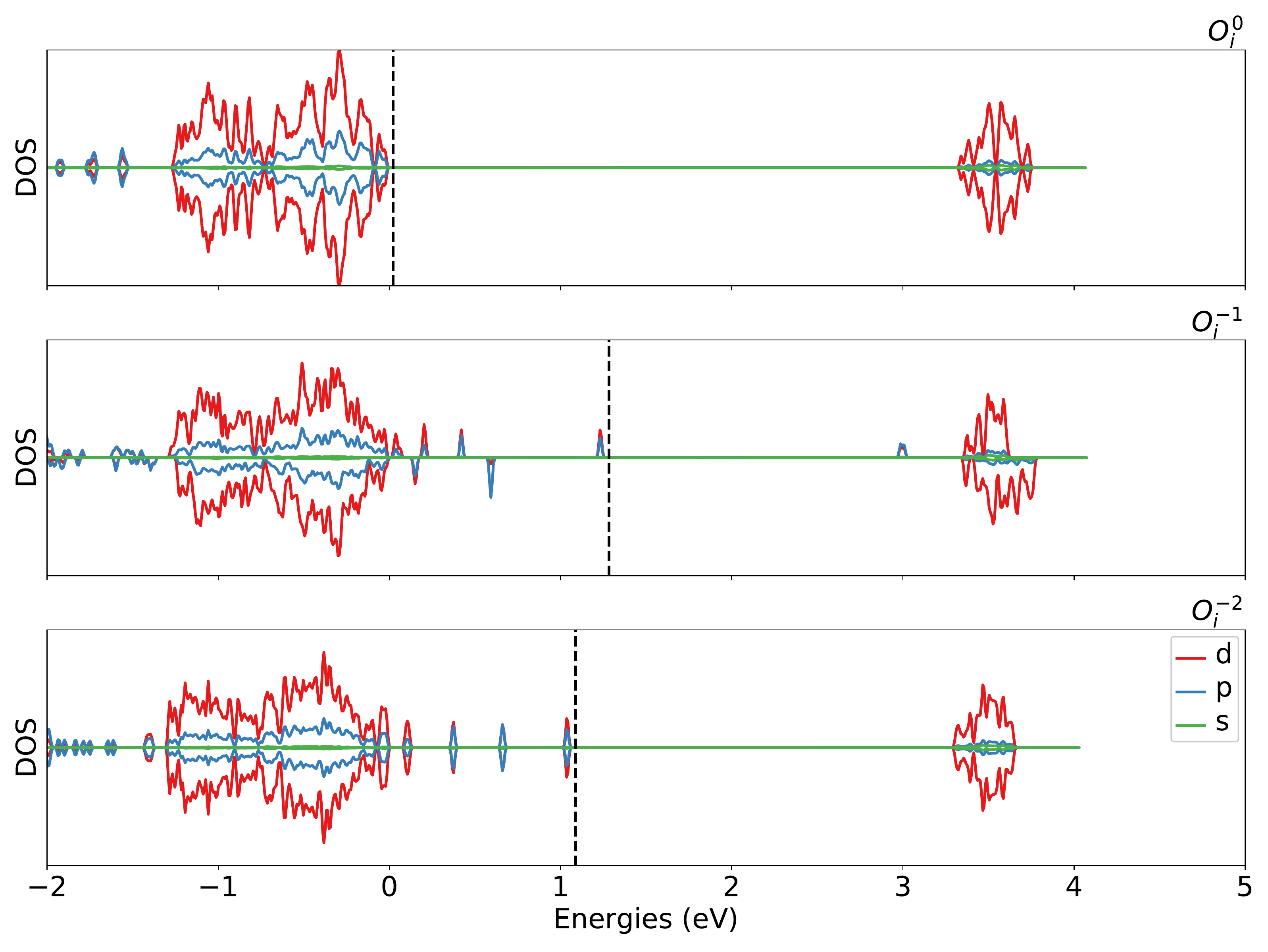}
\caption{O dumbbell interstitials}
\label{fig:O_inter_dos}
\end{subfigure}%
\begin{subfigure}{0.5\linewidth}
\centering
\includegraphics[trim={0.4cm 0.3cm 0 0.4cm},clip,width=1\linewidth]{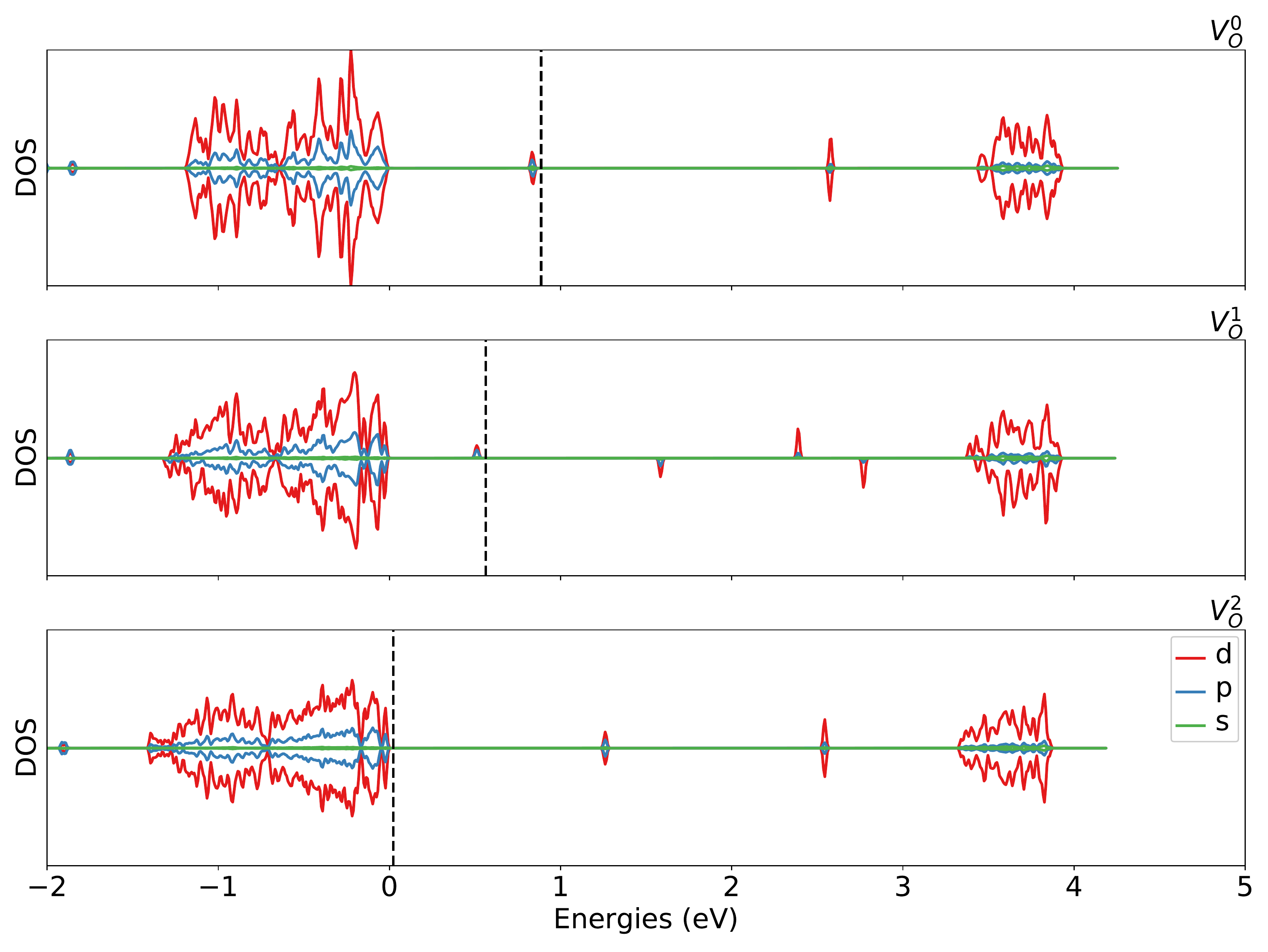}
\caption{O vacancies}
\label{fig:O_vac_dos}
\end{subfigure}
\caption{Orbital projected DOS averaged over all sites for O defects in different charge states. Dashed vertical line marks the position of the Fermi level. The zero level has been fixed at the VBM.}
\label{fig:O_def_dos}
\end{figure*}
Previous studies of defects in \ce{Cr2O3} mainly employed GGA+U method. Given that U and the width of the bandgap often have monotonic relationship U value is often fitted to recover experimental bandgap. The choice of the U-value not only affects the bandgap but also the position of the VBM and to a certain extent CBM with respect to the core bands.   To get an understanding of the influence of U on the defect induced electronic states, we focus on the defect levels of the neutral Cr and O vacancies. In particular, in the studies of Lebreau et al., and Rak and Brenner, the U value of 5.0 eV within Dudarev formalism was used for Cr-3d electrons. According to the DOS of the vacancies reported in these two studies, neutral Cr vacancy exhibits two defects levels, one that is slightly above the VBM (0.2 or 0.3 eV, respectively) , and another that is unoccupied and at the bottom of the CBM. Differing from these studies, Gray et al. used $\{U, J\} = \{4.5, 1\}~eV $ within the Liechtenstein formalism and Medasani et al.  used U=3.7 eV within Dudarev formalism. Both studies indicate that $V_{Cr}$ introduces multiple defect levels. Gray et al.'s results point to four distinct 
defect levels (with two of them very close to each other and slightly below the CBM) in the bandgap, and Medasani et al. reported 5 distinct defect levels. 
If defect levels that are energetically closer are grouped together, then Gray et al and Medasani et al's data predict the existence of three defect level zones: a) just above the VBM, b) 1 eV above the VBM, and c) just below the CBM. The HSE results confirm this pattern (Figure \ref{fig:Cr_vac_dos}). This indicates that $U_{eff}$  of \texttt{\char`\~}3.7 eV  is able to describe defect levels of $V_{Cr}$  more accurately at the GGA+U level of theory. 
For $V_O^0$, HSE predicts two defect levels that are spin degenerate at 0.8 and 2.6 eV above VBM (Figure \ref{fig:O_vac_dos}). Similar electronic structure for $V_O^0$ was obtained in the study of Gray et al., where the spin degenerate levels are at 1.3 and 2.6 eV and in the study of Carey et al, who used $U_{Cr-3d}=5$ eV and $U_{O-2p}=5.5$ eV, where the corresponding defect levels are at 1 and 2 eV above the VBM. Medasani et al. instead observed no degeneracy, 
but the resulting four levels are still within the bandgap. In contrast, Lebreau et al. observed no defect levels within the bandgap for $V_O$. These results indicate that the energetic position of defect electronic states with respect to the VBM are highly dependent on the choice of U parameters and on the 
+U formalism selected.  Comparison with the HSE results shows that although some GGA+U data qualitatively match the HSE results the difference is often quite noticeable especially for neutral vacancies. Interestingly, for charged defects such as $V_{Cr}^{-3}$, $V_O^2$, and for interstitials, defect levels obtained from our GGA+U calculations are in good agreement with those obtained from HSE calculations (see Figures\ref{fig:Cr_vac_dos}-\ref{fig:O_inter_dos}). It is noteworthy  that unlike the study by Gray et al  of native vacancies in \ce{Cr2O3} that assumed fixed defect levels in the bandgap independent of  the defect charge state, our earlier GGA+U and the present HSE studies on vacancies and interstitials indicate that defect levels in the bandgap are not fixed but vary depending on the defect charge state. 

\subsubsection{Charge distribution}
The distribution of the charge of excess hole(s) or electron(s) for the various charge states of the native vacancies and self-interstitials is  shown in SI Fig. S2 - S5.  These plots indicate that both holes and electrons are delocalized. This finding on the delocalization of the holes is consistent with the earlier studies~\cite{kehoe2016,lany_fs,medasani2017}. However, comparison of  charge distributions of $Cr_i^1$ and $Cr_i^2$  holes shows that  delocalization is more pronounced when GGA+U is used.

\subsubsection{0-K Defect Formation Energies}
After structure optimization, the spurious electrostatic interactions inherent in the periodic boundary formalism of charged defects were corrected using the anisotropic FNV (Freysoldt, Neugebauer, and Vande Walle) method~\cite{freysoldt2009,kumagai2014} (see Table S1 in SI for the resulting corrections). The charge transition levels of the defects with respect to the valence band maximum are presented in Table S2 in the SI.
When compared to the corrected transition levels reported in our previous works, the corresponding
transition levels obtained with HSE are overall at a higher Fermi level. 
Some of the transition levels such as $Cr_i$ (3/2) transition, which is at 2.6 eV above the VBM, 
differ by nearly 1 eV (1.64 eV in Ref~\cite{medasani2018}). Further, HSE predicts that only +3 
and +2 charge states are stable for $Cr_i$ within the bandgap. On the other hand, the $O_i$ (0/-2) transition level at 3.07 eV differs only by 0.14 eV from the corresponding transition level 
computed using GGA+U coupled with bandgap correction. This shows that while bandgap 
correction overall improves the position of the  transition levels computed with GGA+U, the 
improved agreement between the HSE computed and badgap corrected GGA+U formation energies is not 
systematic. The position of some transition levels such as $O_i$ (0/-2) and $V_O$ (1/0) are almost the same,
while the position of other transition levels such as $Cr_i$ (3/2) and $V_O$ (2/1) exhibit differences that are greater than 0.4 eV when computed at the HSE and GGA+U  levels of theory.

\begin{wraptable}{l}{0.45\textwidth}
    \caption{Defect formation energies (in eV) of \ce{Cr2O3} with Fermi level ($E_F$) at the middle of the bandgap under O-rich conditions.}
\begin{tabular}{lrcc}
    \hline
    Defect &  q & HSE & GGA+U~\cite{medasani2017,medasani2018} \\
    
    \hline
    \multirow{3}{*}{$V_O$}    &  0 & 5.55 & 6.01 \\
                              & 1 & 5.62 & 6.34 \\
                              & 2 & 6.16 & 7.47 \\
    \hline
    \multirow{4}{*}{$V_{Cr}$}   & 0 & 0.90 & 1.93 \\
                              & -1 & 0.93 & 1.69 \\
                              & -2 & 1.30 & 1.37 \\
                              & -3 & 2.00 & 1.66 \\
    \hline
    \multirow{3}{*}{$O_i$}    &  0 & 2.94 & 2.18 \\
                              & -1 & 4.77 & 3.69 \\
                              & -2 & 5.76 & 4.38 \\
    \hline
    \multirow{4}{*}{$Cr_i$}   & 0 & 11.65 & 10.13 \\
                              & 1 & 9.58 & 8.82 \\
                              & 2 & 7.85 & 7.91 \\
                              & 3 & 6.92 & 7.97 \\
    \hline
\end{tabular}
\label{tab:inter_en}
\end{wraptable}
The dependence of 0K defect formation energies on the position of the Fermi level for both O-rich and Cr-rich conditions shows that not only transition levels, but also the formation energies differ between HSE and GGA+U even after corrections (Figure~S7 in SI). At Cr-rich phase boundary, HSE predicts lower formation energies  than GGA+U for the dominant $Cr_i$ and $V_O$ defects. At $O_2$-rich phase boundary HSE predicted formation energies are lower for neutral Cr vacancies and higher for $O_i$. In contrast, for charged Cr vacancies, the HSE formation energies are in good agreement with those calculated using the GGA+U. These results validate our choice of using HSE to obtain more accurate formation energies needed for the calculation of self-diffusion coefficients

\subsection{Self Diffusion Coefficients}
We evaluated the self-diffusion of Cr and O species mediated by vacancies and interstitials along the diffusion pathways identified in earlier works~\cite{Lebreau2014,medasani2017,medasani2018}. Activation energies for defect migration are sum of  transition state barrier energies computed at the the GGA+U level of theory, as reported in our previous works,~\cite{medasani2017,medasani2018} and the defect formation energies computed at the HSE level of theory. We expect this approach to yield reliable activation energies at lower computational cost due to insensitivity of GGA+U defect migration barrier energies to the variation in U value\cite{rak2018}.

\begin{wrapfigure}{r}{0.43\linewidth}
\centering
\includegraphics[trim={0.4cm 0.3cm 0 1cm},clip,width=1\linewidth]{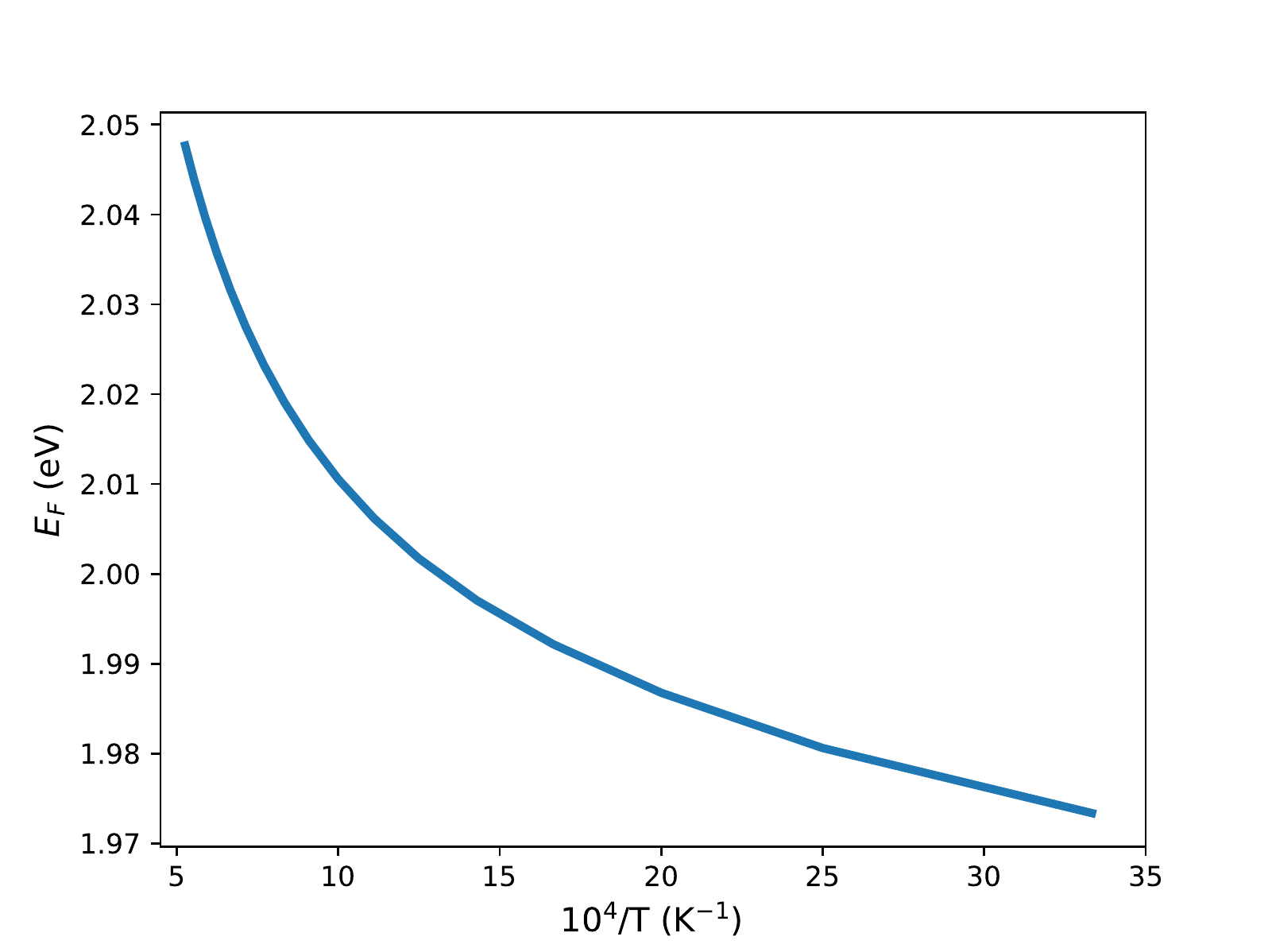}
\caption{ Fermi level in intrinsic \ce{Cr2O3} at Cr/\ce{Cr2O3} interface (Cr-rich condition) as a function of temperature.}
\label{fig:Fermi_interface}
\end{wrapfigure}
\begin{figure*}[!htb]
  \begin{subfigure}{0.5\textwidth}
    \centering
      \includegraphics[trim={0.4cm 0.3cm 0 0.4cm},clip,width=\linewidth]{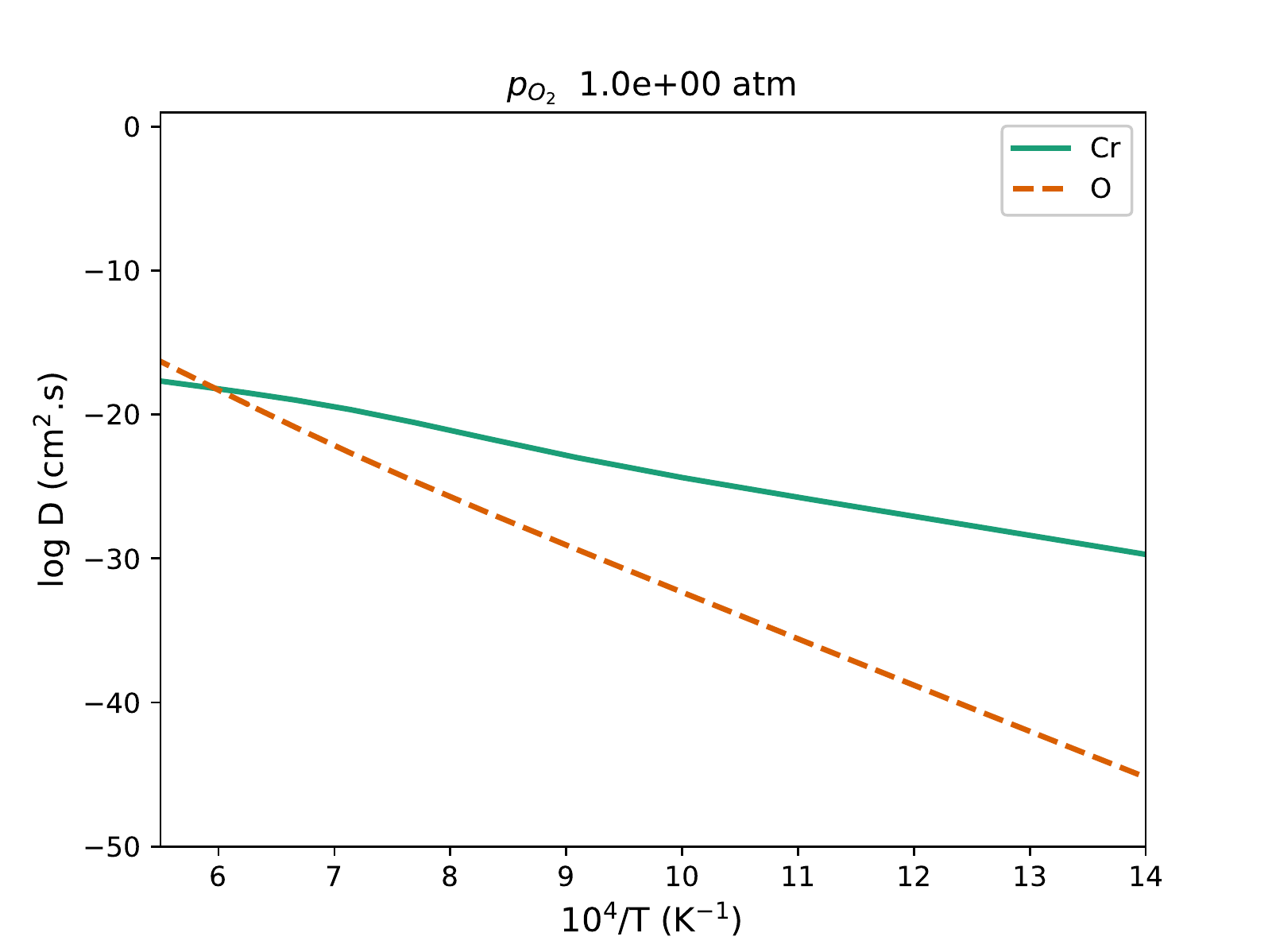}
    \caption{ }
    \label{fig:diff_vs_T_el_P-1}
    \end{subfigure}%
    \begin{subfigure}{0.5\textwidth}
    \centering
      \includegraphics[trim={0.4cm 0.3cm 0 0.4cm},clip,width=\linewidth]{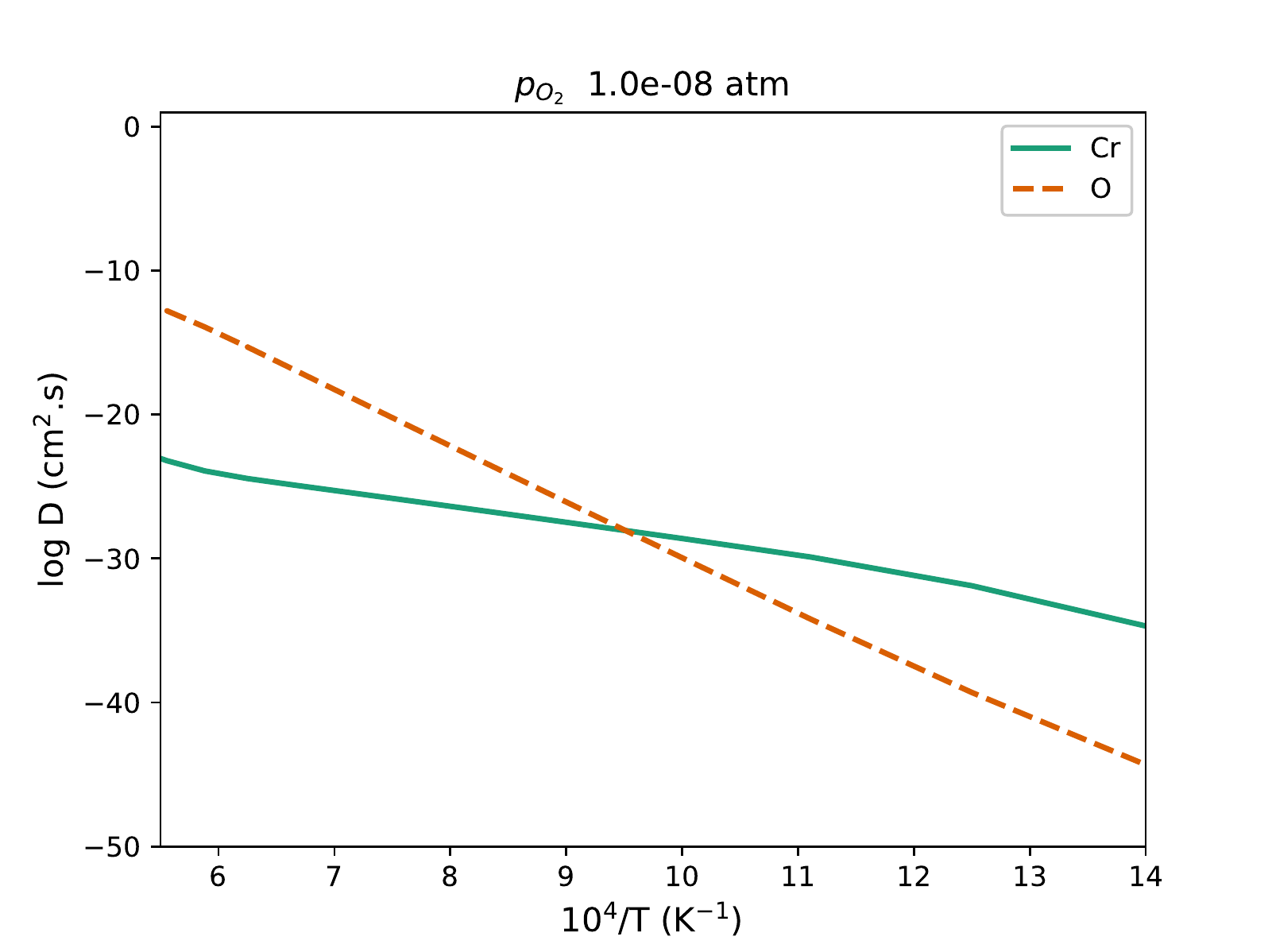}
    \caption{ }
    \label{fig:diff_vs_T_el_P-1e-08}
    \end{subfigure}
  \begin{subfigure}{0.5\textwidth}
    \centering
      \includegraphics[trim={0.4cm 0.3cm 0 0.4cm},clip,width=\linewidth]{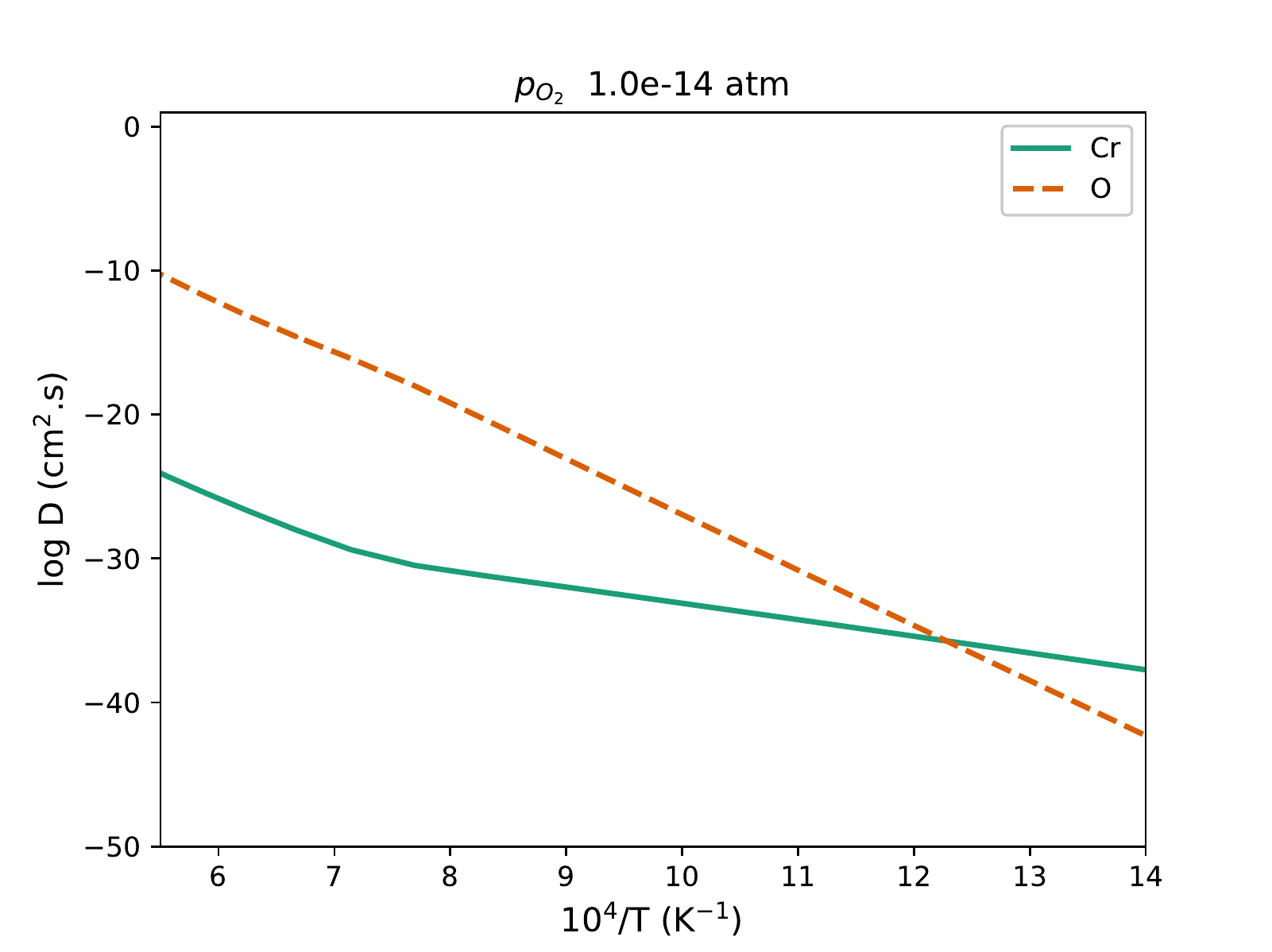}
    \caption{ }
    \label{fig:diff_vs_T_el_P-1e-14}
    \end{subfigure}%
  \begin{subfigure}{0.5\textwidth}
    \centering
      \includegraphics[trim={0.4cm 0.3cm 0 0.4cm},clip,width=\linewidth]{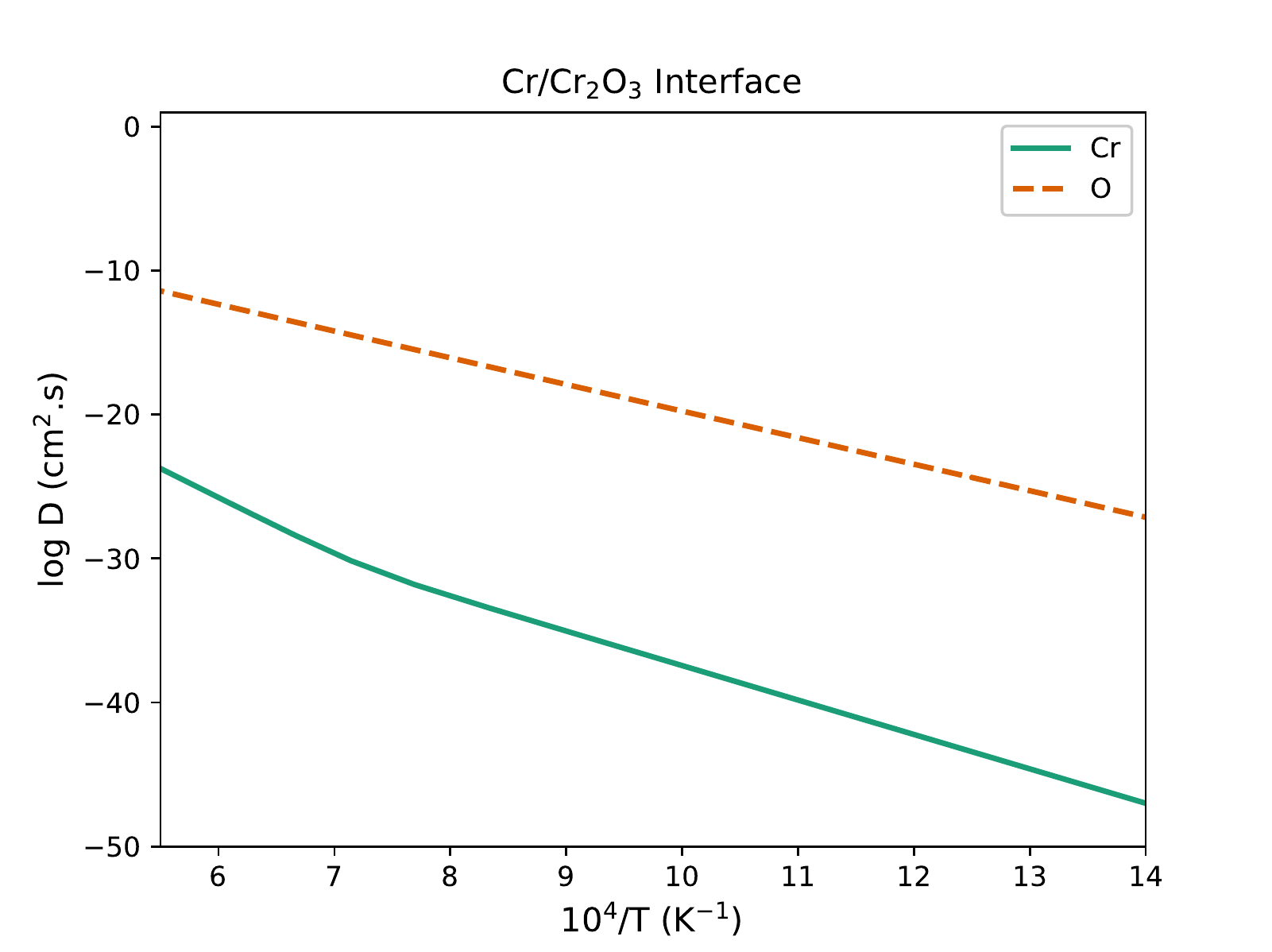}
    \caption{ }
    \label{fig:diff_vs_T_el_interface}
    \end{subfigure}%
  \caption{Self diffusion coefficients of Cr and O at vacuum/oxide interface at \ce{O2} partial pressures of a) 1 atm, b) \num{1e-8} atm, c) \num{1e-14} atm, d) and at Cr/oxide interace.}
\label{fig:diff_vs_T_el}
\end{figure*}

The resulting self-diffusion coefficients are plotted as a function of temperature in Figure~\ref{fig:diff_vs_T_el} at four different conditions: a) standard conditions (1 atm), b) moderate vacuum (\num{1e-8} atm), c) high vacuum (\num{1e-14} atm), and d) metal/metal oxide interface conditions. Electronic effects such as band bending and the  finite electric field arising at the metal-metal oxide and vacuum-metal oxide interfaces are ignored in this study. The plots indicate that at \ce{O2} partial pressure ($p_{O_2}$) of 1 atm, Cr has higher mobility than O. The difference in the mobilities is higher at lower temperatures. As  $p_{O_2}$ is reduced to \num{1e-8} atm,  O become the dominant diffusing species at temperatures higher than 1100 K. At lower temperatures ($<$ 1100 K), Cr still has a higher mobility, but the magnitude of mobility difference is reduced. As $p_{O_2}$ decreases further to \num{1e-14} atm, O mobility becomes higher than that of Cr at even lower temperatures (800 K). At \ce{Cr}/\ce{Cr2O3} interface, relative O mobility is significantly higher than that of Cr at all temperatures studied. 

Sabioni reported the experimental diffusion coefficients of Cr and O as (2.2$\,-\,$7.2)$\times$\num{d-18} \si{\centi\metre\square\per\second}  and (3.2$\,-\,$8.4$)\times$\num{d-18} \si{\centi\metre\square\per\second}  measured  at \ang{1300} C and \ang{1100} C respectively under high vacuum conditions. The experimental diffusion coefficients reported by Sabioni et al. were nearly the same for Cr and O, and relatively independent of $p_{O_2}$. Sabioni proposed the suprising independence of the diffusion coefficients from $p_{O_2}$ could be due to the presence of impurities in the \ce{Cr2O3} samples at around 1 ppm  level. The resulting high concentration of  extrinsic defects may have affected the diffusion coefficients in the \ce{Cr2O3} samples. While the diffusion coefficients reported by Sabioni could not be considered representative of intrinsic \ce{Cr2O3}, they could serve as higher limits on one of the diffusion coefficients by promoting the compensating defects of either Cr or O species. At 1373~K and  at \SI{1d-14}{atm}, which are representative of experimental conditions used to measure  O diffusion, the computed diffusion coefficient of O is found to be at \SI{1d-18}{\centi\metre\square\per\second}. This indicates that O diffusion coefficient was not affected by the extrinsic defects. At 1573~K, the maximal value of Cr diffusion coefficient observed by us is at \SI{1d-18}{\centi\metre\square\per\second} (at standard conditions), which matches with the experimental value of Cr diffusion. 

\begin{wrapfigure}{r}{0.46\linewidth}
  \begin{subfigure}{\linewidth}
    \centering
      \includegraphics[trim={0 0.6cm 0 0.8cm},clip,width=\linewidth]{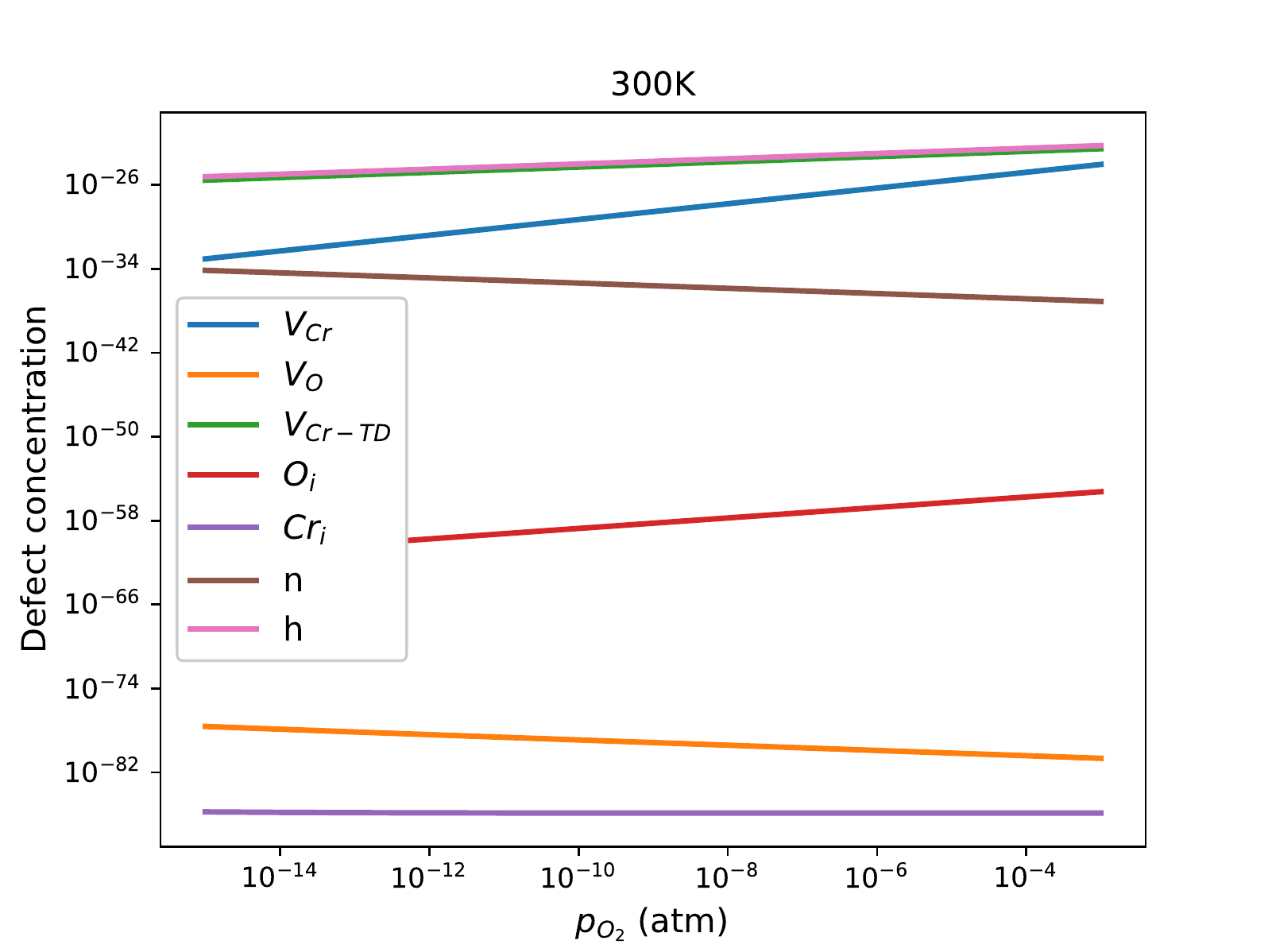}
    \caption{ }
    \label{fig:brouwer_300}
    \end{subfigure}
    \begin{subfigure}{\linewidth}
    \centering
      \includegraphics[trim={0 0.6cm 0 0.8cm},clip,width=\linewidth]{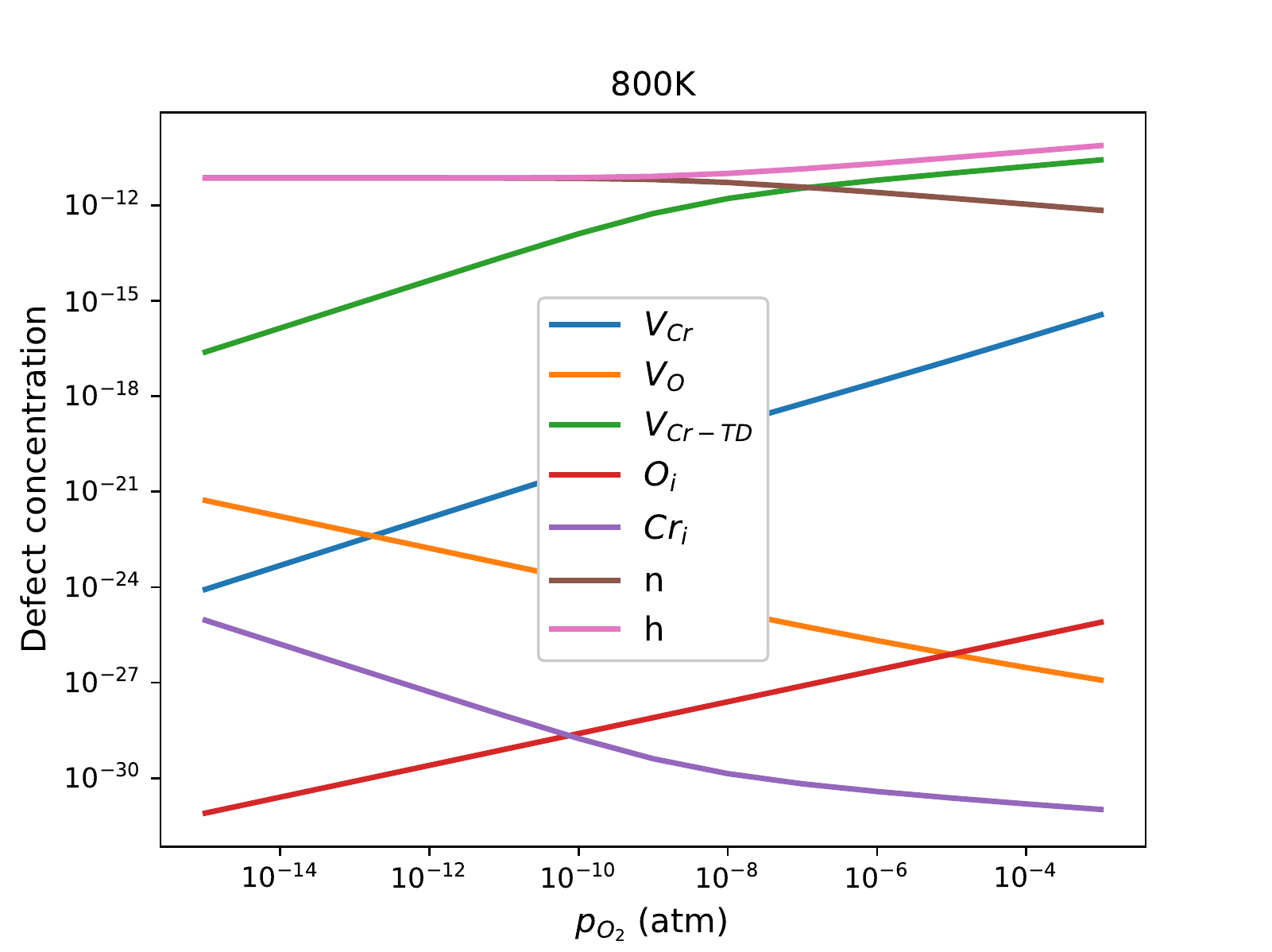}
    \caption{ }
    \label{fig:brouwer_800}
    \end{subfigure}
  \begin{subfigure}{\linewidth}
    \centering
      \includegraphics[trim={0 0.1cm 0 0.8cm},clip,width=\linewidth]{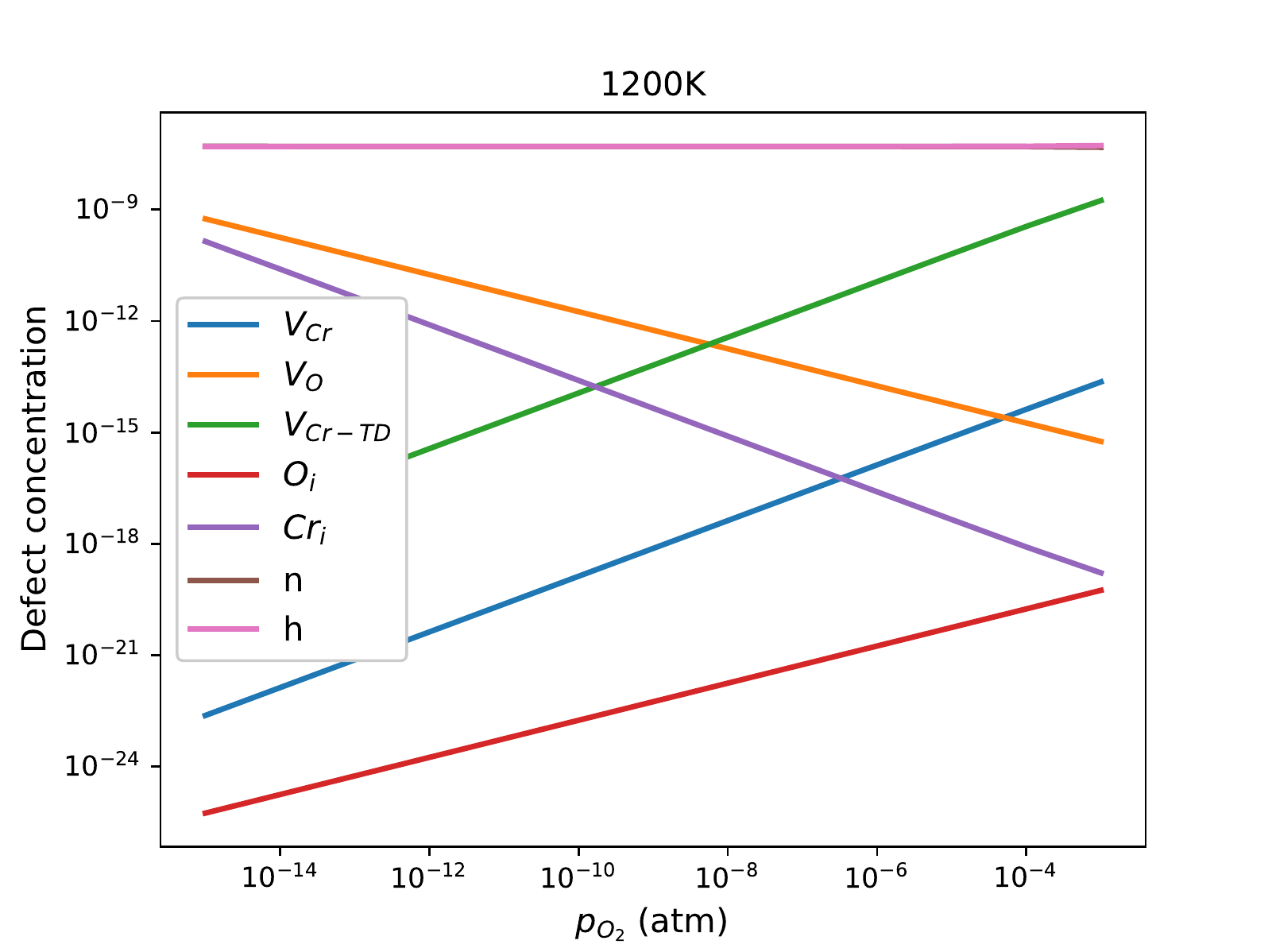}
    \caption{ }
    \label{fig:brower_1200}
    \end{subfigure}%
  \caption{Brouwer diagrams depicting the defect concentrations vs. \ce{O2} partial pressure in \ce{Cr2O3} for a) 300~K, b) 800~K, and c) 1200~K.}
\label{fig:brouwer}
\end{wrapfigure}

To identify the nature of the predominant defects facilitating the diffusion of each species we  plotted the defect concentrations in Brouwer diagrams at different temperatures (Fig.~\ref{fig:brouwer}). At low temperatures ($<$ 800 K), Cr vacancies (including Cr triple defects denoted as $V_{Cr-TD}$) have higher concentrations at all $p_{O_2}$. At 1200 K, the concentration of $V_O$ is higher at $p_{O_2}$ below \num{1e-8} atm.  Above \num{1e-8} atm, Cr vacancies become the dominant defects.  Due to the high concentration of Cr vacancies at pressures close to 1 atm and temperatures above 1200 K, the  self diffusion of Cr is equal to the one measured by Sabioni, suggesting that self-compensating Cr vacancies could be responsible for the high Cr diffusion coefficient  (when compared to the intrinsic Cr diffusion coefficients at corresponding $p_{O_2}$). These results indicate that vacancies are the mediators of reported self-diffusion coefficients in \ce{Cr2O3}.

If the impurity concentration in \ce{Cr2O3} samples could be minimized to very low levels, Fig.~\ref{fig:brower_1200} indicates that at high temperatures ($>$ 1200 K) and low $p_{O_2}$ ($<$ \SI{1d-12}{atm}), $V_O$ and $Cr_i$ could be responsible for O and Cr diffusion respectively. However the relative mobility of O is very high in such cases, indicating again that vacancies dominate the diffusion process in \ce{Cr2O3}.

\begin{figure*}[!htb]
\begin{subfigure}{0.5\textwidth}
\centering
\includegraphics[width=1\linewidth]{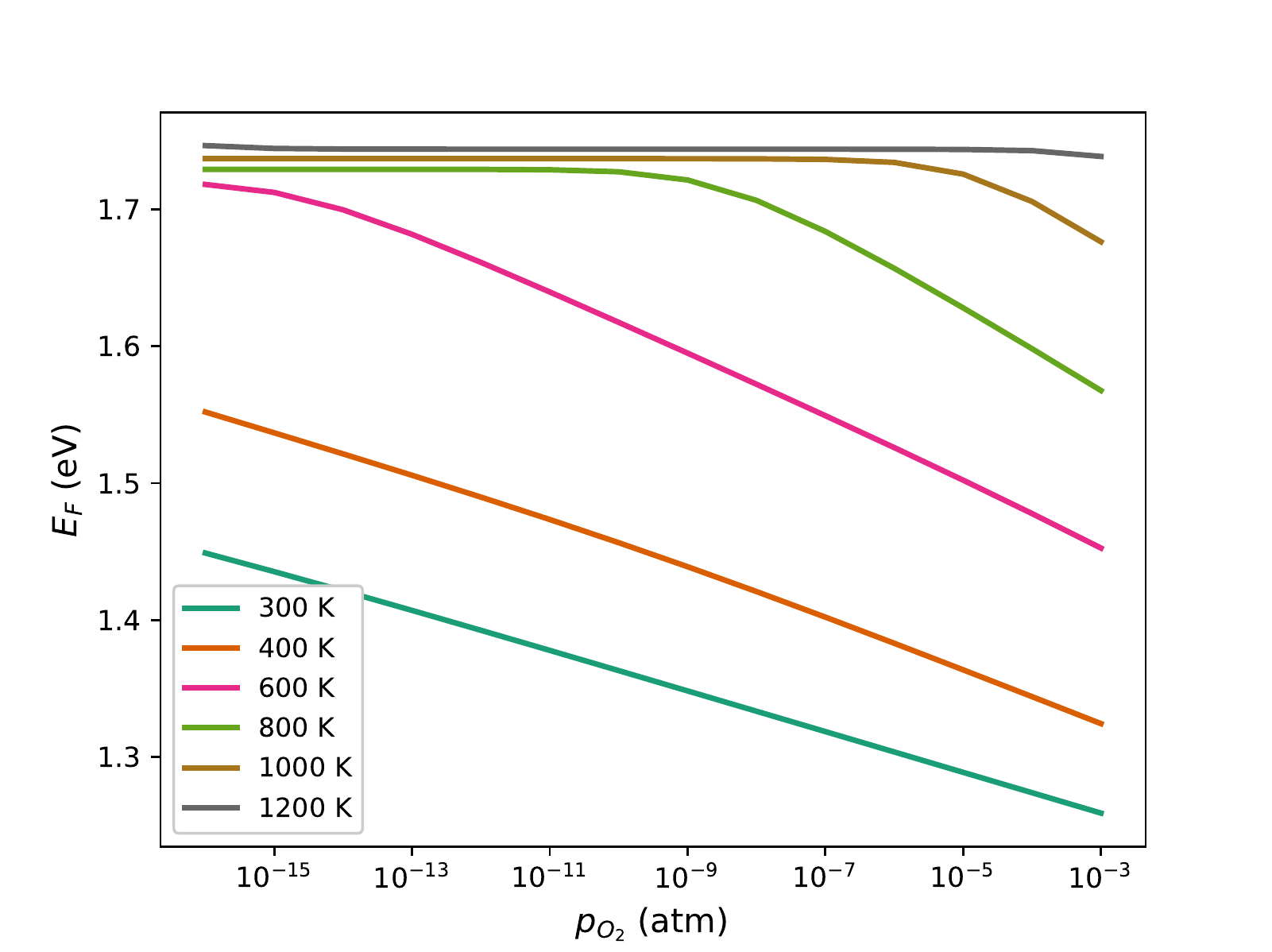}
\caption { }
\label{fig:fermi_variation}
\end{subfigure}%
\begin{subfigure}{0.5\textwidth}
\includegraphics[width=1\linewidth]{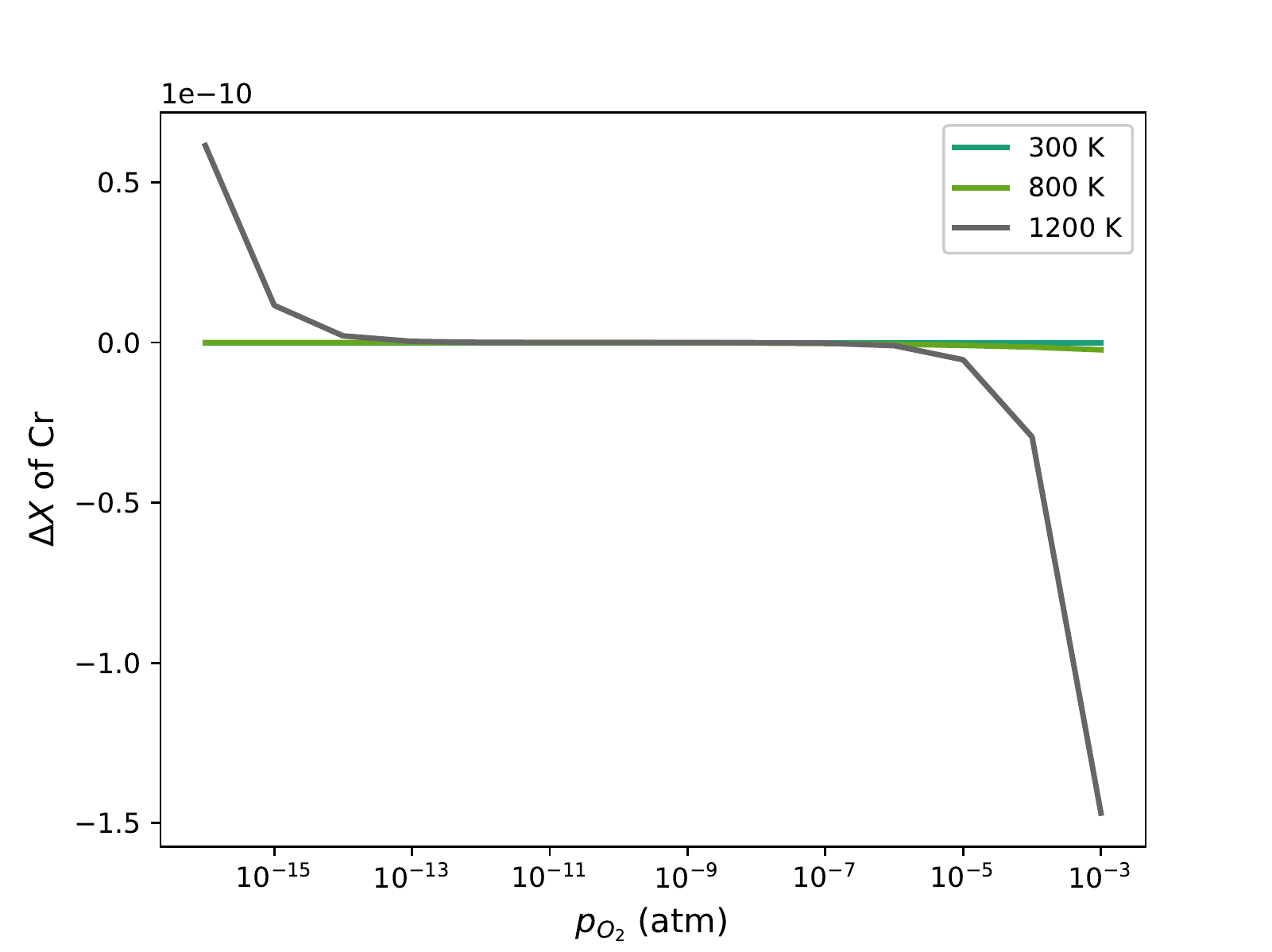}
\label{fig:stoichiometry_variation}
\caption{ }
\end{subfigure}%
\caption{(a) Fermi level in intrinsic \ce{Cr2O3}, and (b) stoichiometry deviation expressed in terms of relative change in Cr composition form stoichiometry as functions of temperature and \ce{O2} partial pressure.}
\label{fig:Fermi_n_stoichoimetry}
\end{figure*}

The Brouwer diagrams also reveal that it would be misleading to use commonly reported 0-K defect formation energies at the phase boundary edges to predict dominant defects responsible for experimentally measured diffusion coefficients. For example at Cr-rich boundary, the 0-K defect energy diagram (Figure S6 in SI) indicates that $Cr_i$ could be the dominant defect if the Fermi level, $E_F$, lies between 1.5 to 1.9 eV. However,  the allowed Fermi level range is only  1.97 $-$ 2.04 eV (Figure~\ref{fig:Fermi_interface}) and neutral vacancies are the dominant defects.  Fermi level variation is plotted as a function of $p_{O_2}$ for different temperatures in Figure~\ref{fig:fermi_variation}. The plots indicate that \ce{Cr2O3} is intrinsically p-type at low temperatures and high $p_{O_2}$, where Fermi level can become as low as 1.2 eV. Under such conditions, the dominant defects are $V_{Cr}$ (Figure S6 in SI). At low temperatures and low  $p_{O_2}$, the Fermi level is around 1.75 eV making \ce{Cr2O3} slightly n-type intrinsically.   Stoichiometry deviation plotted as a function of $p_{O_2}$ in Figure~ \ref{fig:Fermi_n_stoichoimetry}(b) shows that the deviation is negligible in intrinsic \ce{Cr2O3}.

\begin{figure*}[!htb]
  \begin{subfigure}{0.5\textwidth}
    \centering
      \includegraphics[width=\linewidth]{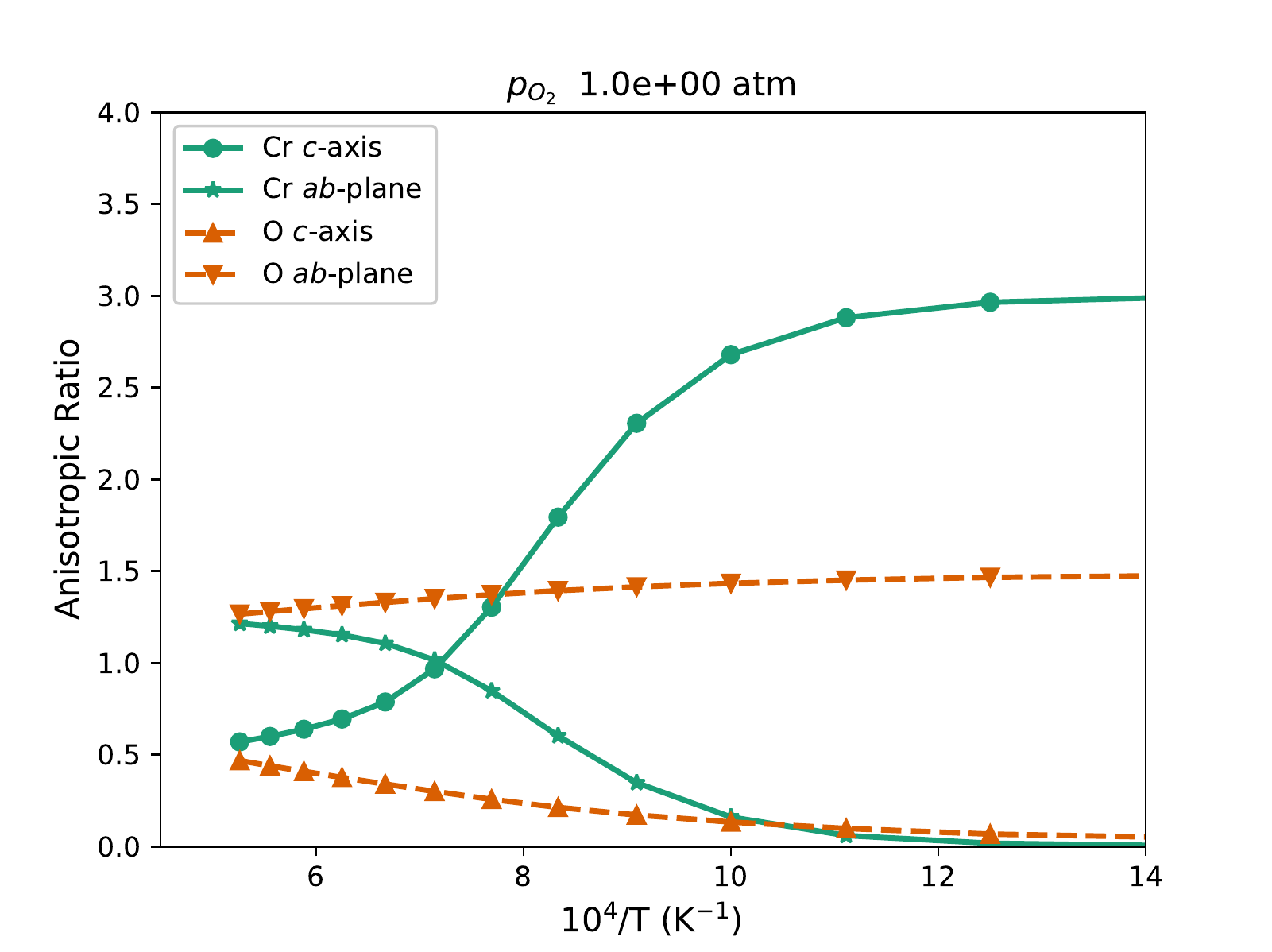}
    \caption{ }
    \label{fig:dir_diff_vs_T_el_P-1}
    \end{subfigure}%
    \begin{subfigure}{0.5\textwidth}
    \centering
      \includegraphics[width=\linewidth]{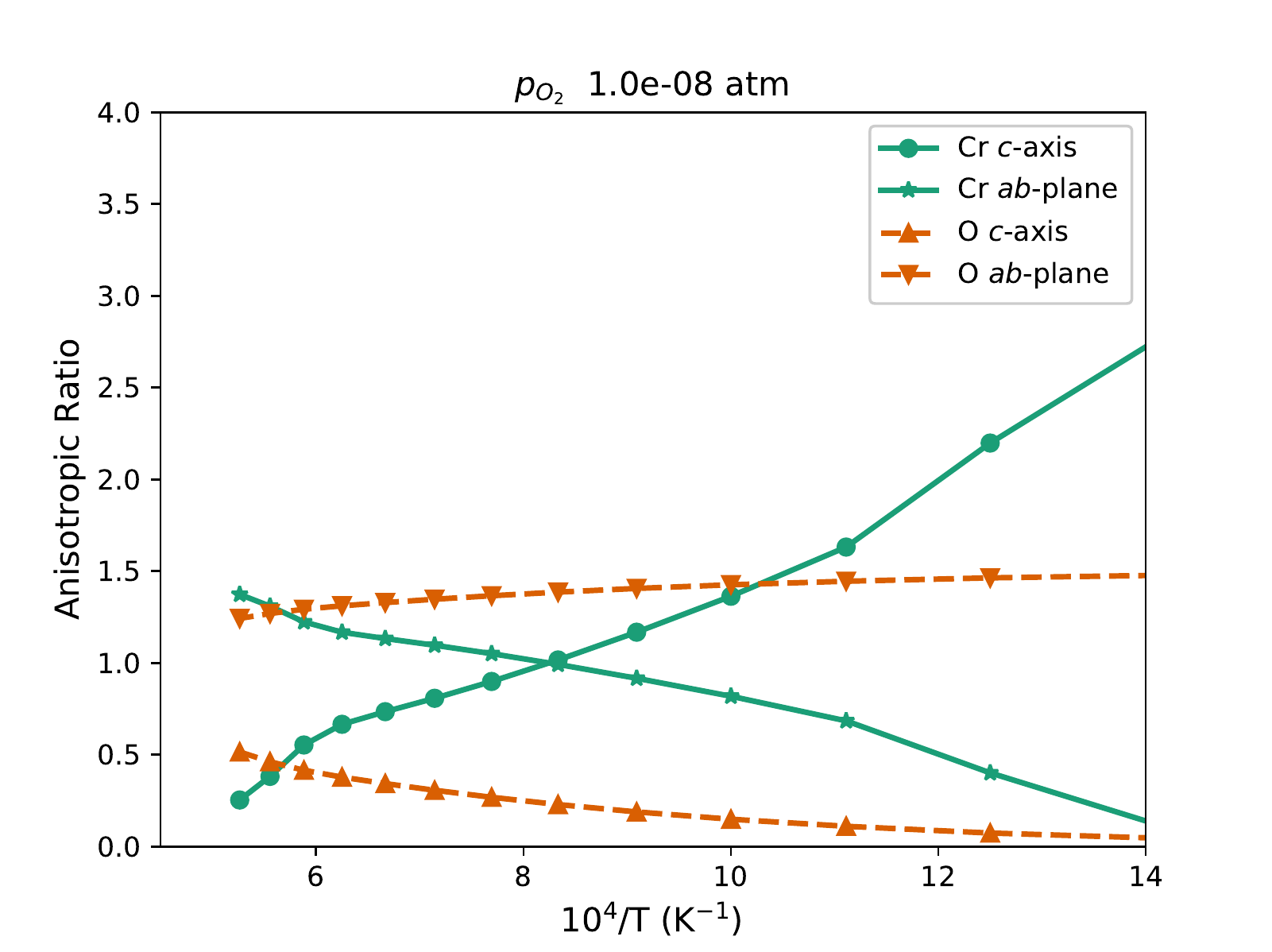}
    \caption{ }
    \label{fig:dir_diff_vs_T_el_P-1e-08}
    \end{subfigure}
  \begin{subfigure}{0.5\textwidth}
    \centering
      \includegraphics[width=\linewidth]{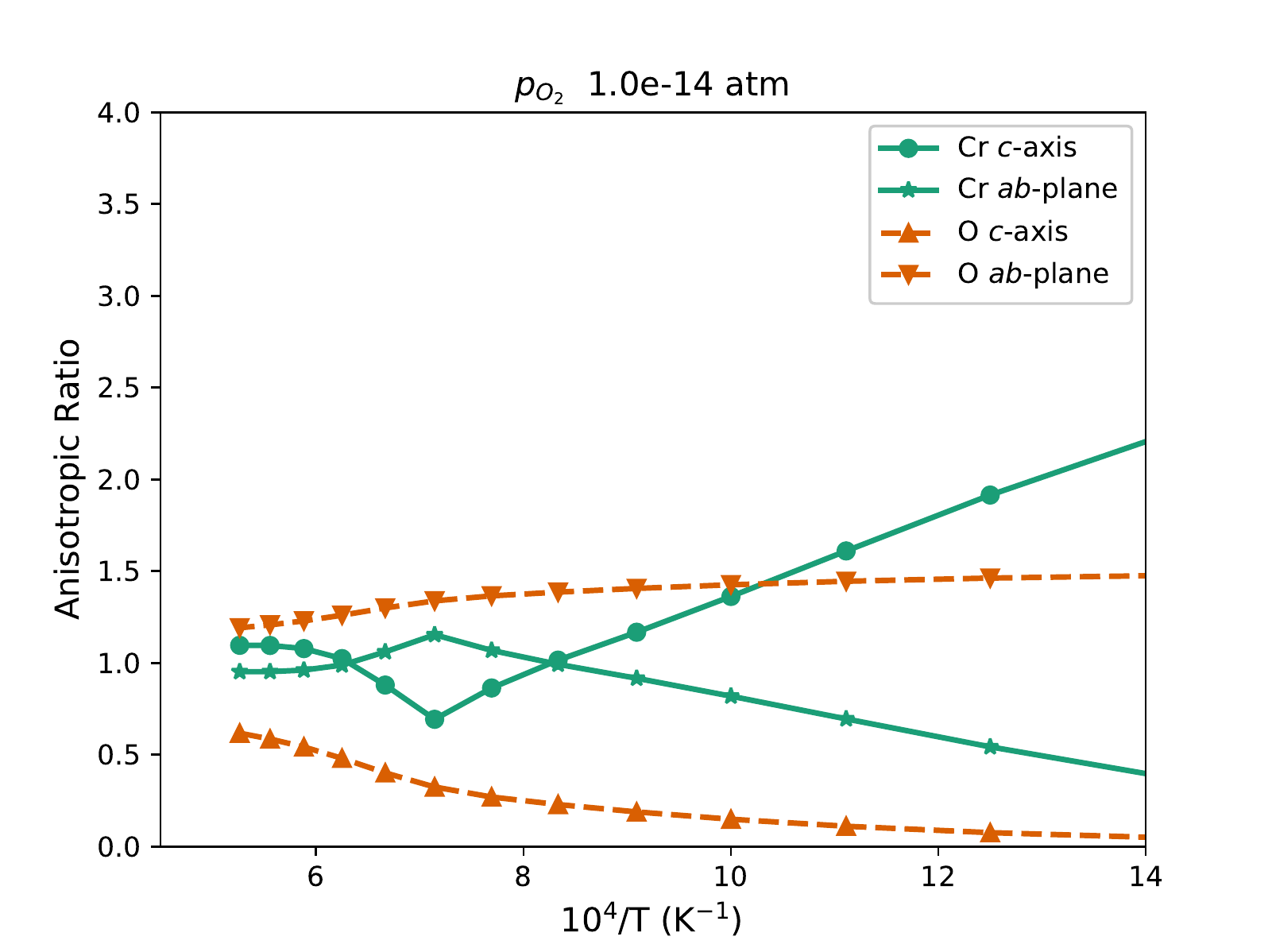}
    \caption{ }
    \label{fig:dir_diff_vs_T_el_P-1e-14}
    \end{subfigure}%
  \caption{Anisotropic ratio of diffusion  of Cr and O at vacuum/oxide interface at \ce{O2} partial pressures of a) 1 atm, b) \num{1e-8} atm, c) \num{1e-14} atm.}
\label{fig:dir_diff_vs_T_el}
\end{figure*}

Figure~\ref{fig:dir_diff_vs_T_el} shows the anisotropic ratio of diffusion for Cr and O at various operational conditions. The results indicate that \ce{Cr2O3} basal plane is the preferential diffusion pathway for O under all conditions. For Cr, however, the preferred diffusion orientation is temperature dependent. At low temperatures, Cr diffuses mainly along the \textit{c}-axis. The mode of diffusion switches to basal plane pathway at around 1200 K under high $p_{O_2}$. In ultra high vacuum conditions of $p_{O_2}\leq$\SI{1d-14}{atm} the primary mode of Cr diffusion switches back to \textit{c}-axis at around 1600 K. Under such conditions, Cr diffusion coefficients along basal plane and \textit{c}-axis become nearly the same.

\section{Summary\label{sec:conclusions}}
Towards a comprehensive  understanding of the diffusion processes in \ce{Cr2O3}, we computed the self diffusion coefficients and Brouwer diagrams in \ce{Cr2O3}. 
Diffusion coefficients were computed  using the DFT at two accuracy levels  and the Einstein random walk formalism.  The defect formation energies were reevaluated with HSE hybrid functional for high accuracy, and  were corrected for spurious electrostatic interaction errors arising in the periodic supercell method. The resulting self diffusion  coefficients have a good agreement with the experimental data. 
Our results reveal that vacancies are primarily responsible for both Cr and O diffusion in intrinsic \ce{Cr2O3}. At high temperatures and low oxygen partial pressures, O has higher mobility. Cr has higher mobility at lower temperatures at moderate to high oxygen partial pressure.  The preferred path for O diffusion is along the basal plane. Cr diffusion takes place along the c-axis at low temperatures and switches to basal plane at temperatures above 1200 K. The revealed mechanism for self-diffusion in \ce{Cr2O3} in a wide range of temperature and pressure conditions opens new avenues for the design of \ce{Cr2O3} materials.

\section*{Acknowledgement}
This work was supported by the U.S. Department of Energy (DOE), Office of Science, Basic Energy Sciences, Materials Sciences and Engineering Division. Simulations were performed using PNNL Institutional Computing facility. PNNL is a multi-program national laboratory operated by Battelle for the U.S. DOE under Contract DEAC05-76RL01830.



\bibliographystyle{elsarticle-num}
\providecommand{\latin}[1]{#1}
\makeatletter
\providecommand{\doi}
  {\begingroup\let\do\@makeother\dospecials
  \catcode`\{=1 \catcode`\}=2 \doi@aux}
\providecommand{\doi@aux}[1]{\endgroup\texttt{#1}}
\makeatother
\providecommand*\mcitethebibliography{\thebibliography}
\csname @ifundefined\endcsname{endmcitethebibliography}
  {\let\endmcitethebibliography\endthebibliography}{}




\end{document}